\global\def\draftcontrol{0}

   \def\versionno{ lines on Calabi-Yau hypersurfaces }

\catcode`\@=11

\expandafter\ifx\csname draftcontrol\endcsname\relax\global\def\draftcontrol{0} 
\fi 

{\count255=\time\divide\count255 by 60 
\xdef\hourmin{\number\count255} 
\multiply\count255 by-60\advance\count255 by\time 
\xdef\hourmin{\hourmin:\ifnum\count255<10 0\fi\the\count255}} 
\def\draftdate{\number\month/\number\day/\number\year\ \ \ \hourmin } 

\newcommand\makepapertitle{\par

  \begingroup 
    \renewcommand\thefootnote{\@fnsymbol\c@footnote}%
    \def\@makefnmark{\rlap{\@textsuperscript{\normalfont\@thefnmark}}}%
    \long\def\@makefntext##1{\parindent 1em\noindent 
            \hb@xt@1.8em{%
                \hss\@textsuperscript{\normalfont\@thefnmark}}##1}%
     \newpage 
     \global\@topnum\z@   
     \@makepapertitle 
     \thispagestyle{empty}\@thanks 
  \endgroup 
  \setcounter{footnote}{0}%
  \global\let\thanks\relax 
  \global\let\makepapertitle\relax 
  \global\let\@makepapertitle\relax 
  \global\let\@thanks\@empty 
  \global\let\@author\@empty 
  \global\let\@date\@empty 
  \global\let\@title\@empty 
  \global\let\title\relax 
  \global\let\author\relax 
  \global\let\date\relax 
  \global\let\and\relax 
  \def\version{\let\version\@version\@gobble} 
} 
\def\@makepapertitle{%
  \newpage 
   \ifnum\draftcontrol=1 {} 
   \version\versionno 
   \vskip 5.5em%
   \else 
   \hfill\hbox to 3.5cm {\parbox{5cm}{\@pubnum}\hss}%
   \vskip 6.5em%
   \fi 
   \begin{center}%
   \let \footnote \thanks 
      {\hskip -0\textwidth \hbox to 1\textwidth%
        {\centerline{\Large\bf{\noindent%
	\parbox[t]{1.3\textwidth}{\begin{center}\@title\end{center}}}}}}%
     \vskip 1.5em%
     {\normalsize
       \lineskip .5em%
       \begin{tabular}[t]{c}%
         \@author 
       \end{tabular}\par}%
     \vskip 1.5em%
     {\@bstract}%
     \end{center}%
     \vfill
     \@date%
     \vskip 1.5em%
   \par 
} 

\gdef\@pubnum{} 
\def\pubnum#1{%
  \gdef\@pubnum{#1}} 

\gdef\@bstract{} 
\def\Abstract#1{%
  \gdef\@bstract{%
   \parbox{\textwidth-0pc}{%
   \centerline{\bf Abstract}\penalty1000 
   \noindent
   \renewcommand\baselinestretch{1.0} 
   {#1}}} 
} 

\gdef\@email{}
\def\email#1{%
   \gdef\@email{%
   Email: {\tt #1}}
}

\def\ps@paper{\let\@mkboth\@gobbletwo%
     \ifnum\draftcontrol=1 
        \def\@oddfoot{\hbox to \textwidth{\tiny \versionno \hfil\tiny\draftdate}%
        \hskip -\textwidth \hbox to \textwidth{\hfil\rm\thepage\hfil}}%
     \else\def\@oddfoot{\hbox to \textwidth{\hfil\rm\thepage\hfil}} 
     \fi 
     \let\@evenfoot\@oddfoot 
} 

\def\body{\clearpage 
          \pagestyle{paper} 
        } 
\newenvironment{acknowledgments}{%
\vskip 3.25ex 
\addcontentsline{toc}{section}{Acknowledgments}
\noindent {\bf Acknowledgments} 
} 

\def\@version#1{\ifnum\draftcontrol=1 
\typeout{}\typeout{#1}\typeout{} 
\vskip3mm\centerline{\hbox{\fbox{\normalsize{\tt DRAFT -- #1 -- } 
                   {\draftdate}}}}\vskip3mm 
\fi} 
\let\version\@version 
\long\def\eqlabel#1{\ifnum\draftcontrol=1 
                    \tag@false  
                    \tag*{(\theequation) \hbox to -0.2cm{\hspace{0cm}\small{#1}\hss}} 
                    \refstepcounter{equation}  
                    \edef\@currentlabel{\theequation} 
                    \ltx@label{#1}          
                    \else 
                    \label{#1} 
                    \fi 
                    } 
\let\st@bibitem\@bibitem 
\let\st@lbibitem\@lbibitem 
\ifnum\draftcontrol=1 
  \def\@bibitem#1{%
    \st@bibitem{#1}\a@@label{#1}\ignorespaces}
  \def\@lbibitem[#1]#2{%
    \st@lbibitem[#1]{#2}\a@@label{#2}\ignorespaces} 
  \def\a@@label#1{%
    \gdef\a@lab{\smash{\normalfont\small#1}} 
    \ifvmode 
      \if@inlabel 
        \global\setbox\@labels\hbox{%
          \llap{\a@lab\let\a@lab\relax 
                \kern\@totalleftmargin\kern\marginparsep}%
          \box\@labels}%
      \fi 
    \fi} 
\fi 

\documentclass[12pt,letterpaper]{article} 

\usepackage{amsmath,bm,amsfonts,amssymb,array,calc,amsthm,rotating}
\usepackage{epsfig,psfrag,graphicx,array} 
\usepackage{graphicx}
\usepackage{color}
\usepackage[colorlinks=true]{hyperref}

\renewcommand\baselinestretch{1.25} 
\renewcommand\section{\@startsection {section}{1}{\z@}%
                                   {-3.5ex \@plus -1ex \@minus -.2ex}%
                                   {2.3ex \@plus.2ex}%
                                   {\normalfont\large\bfseries}} 
\renewcommand\subsection{\@startsection{subsection}{2}{\z@}%
                                   {-3.25ex\@plus -1ex \@minus -.2ex}%
                                   {1.5ex \@plus .2ex}%
                                   {\normalfont\normalsize\bfseries}} 
\renewcommand\subsubsection{\@startsection{subsubsection}{3}{\z@}%
                                   {-3.25ex\@plus -1ex \@minus -.2ex}%
                                   {1.5ex \@plus .2ex}%
                                   {\normalfont\normalsize\it}} 
\renewcommand\paragraph{\@startsection{paragraph}{4}{\z@}%
                                   {-3.25ex\@plus -1ex \@minus -.2ex}%
                                   {1.5ex \@plus .2ex}%
                                   {\normalfont\normalsize\bf}} 
\renewcommand\subparagraph{\@startsection{subparagraph}{5}{\z@}%
                                   {-1.25ex\@plus -1ex \@minus -.2ex}%
                                   {0ex \@plus .2ex}%
                                   {\normalfont\normalsize\it}}

\numberwithin{equation}{section}

\long\def\@makecaption#1#2{%
  \vskip\abovecaptionskip
  \sbox\@tempboxa{{\bf #1:} #2}%
  \ifdim \wd\@tempboxa >\hsize
    {\small\bf #1:} {\small #2}\par
  \else
    \global \@minipagefalse
    \hb@xt@\hsize{\hfil\box\@tempboxa\hfil}%
  \fi
  \vskip\belowcaptionskip}

\renewcommand*\l@section[2]{%
  \ifnum \c@tocdepth >\z@
    \addpenalty\@secpenalty
    \addvspace{.0em \@plus\p@}%
    \setlength\@tempdima{1.5em}%
    \begingroup
      \parindent \z@ \rightskip \@pnumwidth
      \parfillskip -\@pnumwidth
      \leavevmode \bfseries
      \advance\leftskip\@tempdima
      \hskip -\leftskip
      #1\nobreak\hfil \nobreak\hb@xt@\@pnumwidth{\hss #2}\par
    \endgroup
  \fi}
\renewcommand*\l@subsection{\addvspace{-.2em \@plus\p@}\@dottedtocline{2}{1.5em}{2.3em}}
\renewcommand*\l@subsubsection{\addvspace{-.2em \@plus\p@}\@dottedtocline{3}{3.8em}{3.2em}}

\def\hepth#1{\href{http://xxx.arxiv.org/abs/hep-th/#1}{{arXiv:hep-th/#1}}}

\def\math#1{\href{http://xxx.arxiv.org/abs/math/#1}{{arXiv:math/#1}}}

\def\alggeom#1{\href{http://xxx.arxiv.org/abs/alg-geom/#1}{{arXiv:alg-geom/#1}}}
\def\arxiv#1#2{\href{http://xxx.arxiv.org/abs/#1}{{arXiv:#1 [#2]}}}

\definecolor{refcol}{rgb}{0.0,0.0,0.2}
\definecolor{eqcol}{rgb}{.2,0,0}
\definecolor{purple}{cmyk}{0,1,0,0}

\gdef\@citecolor{refcol}
\gdef\@linkcolor{eqcol}
\gdef\@urlcolor{refcol}
\def\colorlinkspurple{\gdef\@urlcolor{purple}}
\def\colorlinksblue{\gdef\@urlcolor{blue}}
\def\colorlinksred{\gdef\@urlcolor{red}}

\def\ie{{\it i.e.}} 
\def\viz{{\it viz.}}
\def\eg{{\it e.g.}} 

\def\cf{{\it cf.}}

\def\revise#1       {\raisebox{-0em}{\rule{3pt}{1em}}%
                     \marginpar{\raisebox{.5em}{\vrule width3pt\ 
                     \vrule width0pt height 0pt depth0.5em 
                     \hbox to 0cm{\hspace{0cm}{%
                     \parbox[t]{4em}{\raggedright\footnotesize{#1}}}\hss}}}}

\def\cala         {{\cal A}} 
 
\def\calc         {{\cal C}}

\def\calh         {{\cal H}} 
 
\def\calj         {{\cal J}} 
 
\def\call         {{\cal L}} 
\def\calm         {{\cal M}} 
\def\caln         {{\cal N}} 
\def\calo         {{\cal O}}

\def\calw         {{\cal W}} 

\def\caly         {{\cal Y}}

\def\complex      {{\mathbb C}} 
 
\def\projective   {{\mathbb P}} 
\def\rationals    {{\mathbb Q}} 
 
\def\zet          {{\mathbb Z}}

\def\del          {\partial} 
 
\def\ee           {{\it e}} 
\def\ii           {{\it i}}

\def\sqr#1#2{{\vcenter{\vbox{\hrule height.#2pt   
 \hbox{\vrule width.#2pt height#1pt \kern#1pt 
 \vrule width.#2pt}\hrule height.#2pt}}}}

\def\Ipp{\mathord{\mathchar "0271 \kern-4.5pt \mathchar"0271}}

\def\zLV {z_{\rm LV}}
\def\zC  {z_{\rm C}}
\def\zG  {z_{\rm G}}
\def\MLV {M_{\rm LV}}
\def\MC  {M_{\rm C}}
\def\MG  {M_{\rm G}}
\def\zD  {z_{\rm D}}
\def\zDo#1{z_{{\rm D}_{#1}}}

\catcode`\@=12 

\begin{document} 


\title{Monodromy of Inhomogeneous Picard-Fuchs Equations}

\date{August 2013}

\author{
Robert A.\ Jefferson$^{\dag}$, Johannes Walcher$^{\dag\ddag}$ \\[0.2cm]
\it $^{\dag}$ Department of Physics, $^{\ddag}$
Department of Mathematics and Statistics \\ 
\it McGill University,
\it Montreal, Quebec, Canada}

\Abstract{
We study low-degree curves on one-parameter Calabi-Yau hypersurfaces, and their
contribution to the space-time superpotential in a superstring compactification 
with D-branes. We identify all lines that are invariant under at least one 
permutation of the homogeneous variables, and calculate the inhomogeneous 
Picard-Fuchs equation. The irrational large volume expansions satisfy the 
recently discovered
algebraic integrality.  The bulk of our work is a careful study of the 
topological integrality of monodromy under navigation around the complex 
structure moduli space. This is a powerful method to recover the single
undetermined integration constant that is itself also of arithmetic 
significance. The examples feature a variety of residue fields,
both abelian and non-abelian extensions of the rationals, thereby providing
a glimpse of the arithmetic D-brane landscape. 
}

\makepapertitle

\body

\version\versionno

\vskip 1em

\tableofcontents

\section{Introduction and Nature of Results}

The reader of this note will appreciate that when it comes to calculations around moduli 
spaces parameterizing supersymmetric vacua of quantum field theories and string theory, 
explicit evaluations of global monodromy rank among both the most subtle and the hardest. 
This is so because except in the simplest situations (really, anything that cannot be 
reduced to the thrice punctured sphere, or hypergeometric functions), the required 
analytic continuations cannot be handled algebraically, and one has to resort to 
numerical methods. (This would be even more true for higher dimensional moduli
spaces.) Moreover, the precise matching of the local data 
from one patch to the next is contingent on keeping track of the chosen continuation 
path, and the relative normalization.

These facts notwithstanding, monodromy calculations are often a worthwhile enterprise.
A priori knowledge of (even part of) the monodromy constitutes valuable information to 
constrain the behaviour around the singular points which are of more direct physical 
and mathematical interest. A posteriori, consistent monodromies
serve as cross-check of local results, and are the final confirmation that all
normalizations are correct. In some situations, such as the one studied in the
present paper, monodromy considerations can be used to determine subtle local data
whose perturbative calculation is either much much harder or even unknown. This
will be the main payoff of the present paper.

Much of the physics motivation for the calculations that we'll present flows from the 
realization that in the context of Calabi-Yau compactifications of type II string theory, 
breaking supersymmetry from eight to four supercharges by wrapping D-branes comes, at 
the level of solving the F-flatness equations on the worldvolume, with an extension of 
moduli spaces, schematically,%
\footnote{We are using a ``mostly mathematical'' notation throughout the paper,
with occasional physics terminology when missing words. Eq.\ \eqref{extension} means
that there exists a nice map from the $\caln=1$ moduli space to the $\caln=2$ moduli
space, as we presently explain. As physicists, we would point the arrow in the
direction of lower supersymmetry, as in, $\caln=2\to\caln=1$.}
\begin{equation}
\eqlabel{extension}
\calm_{\caln=1}\longrightarrow \calm_{\caln=2} 
\end{equation}
where the fiber of the map are the $\caln=1$ ``open string'' vacua with fixed value 
of $\caln=2$ ``closed string'' moduli. In a fixed charge sector, the extension is 
finite modulo continuous open string moduli, and is accompanied with a rich 
algebraic structure whose physical consequences are only beginning to emerge. The 
aspect emphasized in \cite{arithmetic} is the action of the Galois group on the
extending vacua, locally around large volume point. What we study in the present
paper is how these local extensions fit together into the global structure of 
\eqref{extension}. This is a generalization of the work \cite{lawa}.

In the rest of the introduction, we describe the geometric (Hodge theoretic)
situation, and then 
summarize our main results. The bulk of the paper is devoted to explicit 
calculations. We include a brief discussion section at the end, but the broader 
lessons for the landscape of $\caln=1$ string vacua will be extracted elsewhere.

\medskip

The geometric situation underlying our calculations involves, first of all, a 
smooth, quasi-projective family of Calabi-Yau threefolds $\caly\to B$, with
semi-stable compactification $\bar\caly\to\bar B$. To keep that part simple,
we'll be working with the earliest list of four examples, one-parameter hypersurfaces
in weighted projective space, originally studied in \cite{morrison,klemmtheisen}.
The list includes (the mirror manifolds of): the quintic $\projective^4_{11111}[5]$,
the sextic $\projective^4_{11112}[6]$, the octic $\projective^4_{11114}[8]$,
and the dectic $\projective^4_{11125}[10]$. So, the base of our family will be
a thrice-punctured projective line, which we parameterize with a complex variable
$z$ taking values $0,1,\infty$ at the three singular points:
$B\cong\projective^1\setminus\{0,1,\infty\}$. 

The middle cohomology groups of the members of our family, $H^3(Y_z,\complex)$,
are 4-dimensional symplectic vector spaces, and, as $z$ varies over $B$, fit 
together to a holomorphic vector bundle, $\calh_\complex$, that is naturally flat because
the fibers contain the locally constant integral lattice $H^3(Y_z,\zet)$, fitting together
to the local system $\calh_\zet$. The global structure of the bundle $\calh_\complex=
\calh_\zet\otimes\calo_B$ is encoded in the monodromy representation
\begin{equation}
\eqlabel{monodromy}
\rho : \pi_1(B) \to {\rm Sp}(4,\zet)
\end{equation}
In the usual conventions, the boundary point $z=\zLV=0$ is the point of maximal 
unipotent monodromy (large volume point), $z=\zC=1$ is the conifold point with
unipotent monodromy of rank 1, while the Gepner point $z=\zG=\infty$ has 
monodromy of finite order ($5$, $6$, $8$ and $10$ in the four examples, 
respectively). We'll imagine the base point implicit in \eqref{monodromy} to be
located close to $\zLV$, and denote the corresponding fixed symplectic lattice 
by $(H_\zet,\langle\cdot,\cdot\rangle)$. 
A choice of basis in $H_\zet$ allows to write matrices representing
generators of $\pi_1(B)$, which we denote by $\MLV$, $\MC$, $\MG$, and which
satisfy
\begin{equation}
\MLV\cdot\MC = \MG
\end{equation}

A specific basis of $H_\zet$ is determined from the intrinsic properties of the 
variation of Hodge structure associated with $\caly\to B$, and its degeneration at
$\bar B\setminus B$. From the fact that 
$N:=\MLV-{\it id}$ is nil-potent ($N^4=0$) and maps integral vectors to integral 
vectors, the spaces ${\rm Im}(N^{3-j})/{\rm Im}(N^{4-j})$ for $j=0,1,2,3$ are 
(projectively) rational. Using additional information from the symplectic form
determines an integral basis $(\gamma_0,\gamma_1,\gamma_2,\gamma_3)$ of $H_\zet$ 
that is adapted to the monodromy weight filtration
\begin{equation}
\eqlabel{monoweight}
W_{2j} = {\rm Ker} N^{j+1}\,,\qquad j=0,1,2,3
\end{equation}
(namely $\gamma_j$ generates $W_{2j}/W_{2j-2}$), up to a lower-triangular symplectic
transformation that is integral except for a single constant of integration, $\alpha$,
which corresponds to an indeterminacy $\gamma_3\to\gamma_3+\alpha\gamma_0$, and
can a priori take any (imaginary) value. These constants can be determined
and hence a basis fully fixed by studying the behaviour at the
conifold locus. Namely, one imposes that $\gamma_3$ vanish at $z=\zC$, and 
$\MC$ send $\gamma_0\mapsto\gamma_0+\gamma_3$, and leave $\gamma_1$, $\gamma_2$, 
untouched.

In practice, the task is accomplished by calculating the periods of the holomorphic
three-form, \viz, the restrictions
\begin{equation}
\eqlabel{viz}
\varpi_j = \langle{\gamma_j},\cdot\rangle\bigr|_{F^3\calh}
\end{equation}
via the polarization $\langle\cdot,\cdot\rangle$, to the first step of the Hodge 
filtration $F^*\calh$ on $\calh_\complex$. More precisely,
choosing a non-zero section $\Omega\in\Gamma(B,F^3\calh)$ determines a Picard-Fuchs 
differential equation satisfied by any (complex) period
\begin{equation}
\call\varpi(\Omega) = 0\,,
\end{equation}
and studying the analytic properties of the solutions of that differential equation
provides all the data listed above. The first such calculation was completed for
the quintic in \cite{cdgp}, with further explanations in \cite{morrison2,deligne}.
For more recent discussions, see \eg, \cite{doranmorgan} which in particular points 
out a severe ambiguity of this procedure, or \cite{ggk1}.

Once this is done, the (conjecturally) irrational constant $\alpha$ mentioned above 
features as an entry of the limiting period matrix with respect to the canonical mirror 
map coordinate,
\begin{equation}
\eqlabel{canonical}
q = \exp 2\pi\ii \frac{\langle\gamma_1,\cdot\rangle}{\langle\gamma_0,\cdot\rangle}
\biggr|_{(F^3\calh)^\times}
\end{equation}
The fact that in general,
\begin{equation}
\eqlabel{ingeneral}
\alpha\in \frac{\zeta(3)}{(2\pi\ii)^3} \rationals
\end{equation}
is explained by the physics origin of this constant (in perturbative corrections to the
sigma-model on the A-model manifold), as well as by motivation \cite{ggk1}. 
It also meshes nicely with the recent discussions of integral structures on quantum 
cohomology in the context of the gamma genus, see \cite{KKP,maxim,mattchuck}.

\medskip

We now introduce the main complication, which models extension by D-branes \eqref{extension}. 
It is the same as in several previous works \cite{mowa,krwa,newissues}. For each fixed
member of our family of Calabi-Yau threefolds, we find holomorphic curves $C_{z,k}\subset 
Y_z$ that vary generically locally uniquely (as algebraic cycles modulo algebraic 
equivalence) with 
$z$. Here, $k$ is an index running over a certain finite set $A$. Consideration of such a 
finite collection of curves is necessary because any given $C_{z,k}$ will, under 
continuous global variation of $z$, branch at specific locations in $B$, \ie, the local
variation will not be unique, and the curve will not return to itself when the variation
encircles those branch points. In other words, in order for the collection of curves to
fit together into a globally well-defined algebraic cycle $\calc$, we first have to extend 
the moduli space to an $|A|$-fold branched covering $\hat B\to B$. Schematically,
\begin{equation}
\eqlabel{scheme}
\begin{array}{ccccc}
\calc&\subset&\hat\caly &\longrightarrow& \caly \\
~\downarrow~ & & ~\downarrow~ & & ~\downarrow~\\
\hat B &=\!=& \hat B &\longrightarrow & B 
\end{array}
\end{equation}
and the $C_{z,k}$ are components of the fibers of $\calc\to B$. We assume that for
fixed $z$, the $C_{z,k}$ for different $k$ are homologous to each other, and generically
irreducible. We'll call the branch locus of $\hat B\to B$ 
the ``open string discriminant'', and denote it by $D$. In the examples, $D$ 
is a finite number of points. (To be sure, the extension \eqref{scheme} extends 
to the compactification $\bar\caly \to\bar B$, and $\hat{\bar B}\to\bar B$ can also 
be branched at $\bar B\setminus B$. This plays an important role in our analysis.
But when we speak of open string discriminant, we only mean points that were not
on the boundary before.)

Associated to the algebraic cycle $\calc\to B$, we have a variation of mixed Hodge 
structure. Locally on $B$, the extension is encoded in the Abel-Jacobi map to the 
intermediate Jacobian,
\begin{equation}
\calj = F^2\calh \!\setminus\!\calh_\complex / \calh_\zet
\end{equation}
as discussed extensively in the literature, {\it loc.\ cit.}. 
Specifically, to a local family of
homologically trivial cycles, such as $C_z=C_{z,k}-C_{z,k'}$ in some simply connected
open set in $B$, we can associate a normal function, $\nu$, as a holomorphic section 
of $\calj$ satisfying Griffiths transversality
\begin{equation}
\eqlabel{trans}
\nabla\tilde\nu \in F^1\calh\otimes \Omega_B
\end{equation}
Here $\tilde\nu$ is a lift of $\nu$ to $\calh_\complex$ and $\nabla$ is the Gauss-Manin 
connection. Such a lift can be conveniently represented by an integral over a
three-chain bounding $C_z$. Actually, as explained in \cite{mowa} as a consequence of
\eqref{trans} (and surjectivity of the infinitesimal period mapping), the {\it complete} 
information about the extension class $\nu$ can be recovered from the inhomogeneous 
Picard-Fuchs equation
\begin{equation}
\eqlabel{iPF}
\call \tau(\Omega) = f
\end{equation}
satisfied by the truncated normal function
\begin{equation}
\eqlabel{truncated}
\tau = \langle\tilde\nu,\cdot\rangle|_{F^3\calh}
\end{equation}
(\cf, \eqref{viz}). Moreover, as emphasized in \cite{newissues}, the inhomogeneity
on the right hand side of \eqref{iPF} is local and additive in the boundary
cycle $C_z$. Since the integral (period) ambiguity of $\tau$ drops out of the 
differential equation \eqref{iPF}, this means that we can associate an 
inhomogeneity, $f_k$, to each curve, $C_{z,k}$, by itself, such that when 
$C_z=C_{z,k}-C_{z,k'}$, we have
\begin{equation}
f = f_k - f_{k'}
\end{equation}
More formally, and for the global issues which we propose to study in the present 
paper, it is convenient to fix a ``marking'' on $A$, the finite set labeling
the $C_{z,k}$. The simplest way to do this is to include a locally constant 
(technically, of vanishing infinitesimal invariant) and globally invariant curve 
in the same homology class that serves as reference point for the chain integrals. 
In the calculations, this additional cycle will often be implicit, though we promise 
to display it at least once (see eq.\ \eqref{promise}).

Physically, in an $\caln=1$ compactification of the type II/I superstring, the 
truncated normal function $\tau$ gives the contribution to the space-time 
superpotential $\calw$ for the chiral scalar fields coming from $\caln=2$ 
vector-multiplets, that is made by a D-brane configuration whose algebraic 
characteristic class is the cycle under consideration, after integrating out all 
(massive) degrees of freedom on the D-brane worldvolume.

To state the main results of our calculations, we denote by $\cala$ the
local system obtained by tensoring the data of the extension $\hat B\to B$
with $\zet$. Continuation of the bounding chains over $B$ then really is an
extension of local systems
\begin{equation}
\calh_\zet\longrightarrow \hat\calh_\zet \longrightarrow \cala
\end{equation}
that underlies the variation of mixed Hodge structure, and which we 
recover from the solutions of the inhomogeneous Picard-Fuchs equation. In
other words, we will determine the monodromy representation
\begin{equation}
\eqlabel{underlies}
\hat\rho:\pi_1(B\setminus D) \to S_A\times i{\rm Sp}(4,\zet)
\end{equation}
where $S_A$ is the symmetric group and
\begin{equation}
i{\rm Sp}(4,\zet) = (H_\zet)^A \ltimes {\rm Sp}(4,\zet)
\end{equation}
The factor $(H_\zet)^A$ arises because the bounding chains will only
return up to closed three-cycles, and manifests itself in shifts of the
truncated normal function by solutions of the homogeneous equation. The
crux of the computation is that these shifts are indeed integral periods. 

We find that, in analogy with the homogeneous case reviewed above, the integrality 
of monodromy can be determined by combining data from the large volume point and 
the conifold. This data can be interpreted in terms of limiting values of normal functions
studied in full generality in the work of Green-Griffiths-Kerr \cite{ggk2}. 
In a degeneration of maximal unipotent monodromy, the for us relevant statement 
is that with respect to the monodromy weight filtration \eqref{monoweight}, the 
lift of the normal function is integral modulo $W_1$, and rational modulo $W_0$. 
(It is integral modulo $W_0$ when the covering $\hat{\bar B}\to\bar B$ is trivial at 
$\zLV$.) Moreover, the coefficient $a_k$ of the fundamental period $\varpi_0$ in the 
truncated normal function $\tau_k$ (in the limit $z\to\zLV$, with respect to
the canonical coordinates \eqref{canonical}) has an interpretation 
in terms of the geometry of the singular fiber $Y_{\zLV}$, which leads to general 
expectations about the range of values analogous to \eqref{ingeneral}. In our
examples, we find 
that this coefficient is completely determined from the conifold monodromy, and 
our numerical results are consistent with the general expectations.

An interesting observation is that, at least in all examples that we study, the
integral structure at $\zLV$, (and in particular, the constants $a_k$) can {\it also}
be determined by tracking the vanishing normal function to the open string 
discriminant, $D$, and imposing appropriate boundary conditions over there. We remark
that this possibility is not a priori obvious (at least to us) because it 
requires a certain relation between the branch structure at $\zLV$ and the 
number of components of $D$. We will emphasize this aspect in the discussion. 
As a practical matter, however, the coincidence is quite welcome because it 
over-constrains integrality of monodromy.

\medskip

Here is an overview over the remainder of the paper:

\medskip

We'll start in section \ref{curves} by identifying interesting cycles
$\calc$ in each of our four families of Calabi-Yau hypersurfaces. As in 
\cite{arithmetic}, we organize the search by looking for curves of low degree,
and lines specifically. Drawing on the strategy employed by van Geemen
\cite{albanokatz,mustatathesis,mustata}, imposing certain discrete symmetries 
allows us to fully solve the problem in certain cases. We note that some of our
lines actually belong to families (in the sense that they allow additional 
continuous deformations for fixed $z$), but we do not complete the analogue 
of the discussion of van Geemen lines in \cite{mustatathesis,cdvv}. Referring the 
interested reader to \cite{mowa,newissues} for the details of the method, 
and to appendix \ref{few} for a few intermediate steps in one example, 
we present the result of the calculation of the inhomogeneity $f_k$ for each 
of our cycles.

In section \ref{algebraic}, we localize our cycles to the large volume point 
$\zLV$. Following \cite{arithmetic}, we perform a Newton-Puiseux expansion
that separates the curves by residue field. We then check that the A-model 
expansion of the truncated normal function (the space-time superpotential) 
satisfies the ``D-logarithm integrality'' discovered in \cite{arithmetic}, 
and recently proven in \cite{svw}. All the new cycles from section \ref{curves} 
turn out to have residue fields that are abelian extension of $\rationals$. 
Therefore, in order to have a more complete set of examples for the monodromy
calculations, we also include the (non-abelian) conics from \cite{arithmetic}.

Section \ref{topological} then is concerned with the main calculations. For
each of the cycles, we expand periods and truncated normal functions at the
conifold and at the open string discriminant. Numerical analytic continuation
along certain paths in $B$ determines the relevant change of basis and monodromy
matrices. The one friendly aspect is that all components of the open string 
discriminants are on the real axis.

We summarize our numerical results in section \ref{discussion}, see in particular 
table \ref{main}, and discuss the arithmetic significance to the best of our
abilities.

\section{From Curves to Residues}
\label{curves}

For completeness, we begin with some of the standard homogeneous data. Our Calabi-Yau
manifolds are hypersurfaces of degree $d=5,6,8,10$ in weighted projective space,
\begin{equation}
\{ W = 0 \}  \subset \projective^4_{w_1,w_2,w_3,w_4,w_5}
\end{equation}
where $(w_1,w_2,w_3,w_4,w_5)=(1,1,1,1,1),(1,1,1,1,2),(1,1,1,1,4),(1,1,1,2,5)$
are the weights, and $d=\sum_i w_i$. 
The Fermat-polyhedron-Dwork pencil from which we construct the mirror manifold
is specified by the family of polynomials
\begin{equation}
W = \sum_i \frac {w_i}{d} x_i^{d/w_i} - \psi \prod_i x_i
\end{equation}
where the global complex structure parameter is related to $\psi$ via
\begin{equation}
z = \psi^{-d} 
\end{equation}
The convenient normalization of the holomorphic three-form is
\begin{equation}
\eqlabel{convenient}
\Omega = \frac{|G|}{(2\pi\ii)^3} {\rm Res}_{W=0} \frac{\psi\omega}{W}
\end{equation}
where $\omega=\alpha(v)$, $\alpha=dx_1\wedge\ldots\wedge dx_5$, 
$v=\sum w_i x_i\partial_i$, and $|G|= d^3/\prod w_i $ is the order of
the Greene-Plesser group. The three-form satisfies the Picard-Fuchs equation
\begin{equation}
\eqlabel{where}
\call \Omega = d\beta
\end{equation}
where $\beta$ is a certain two-form. The Picard-Fuchs operator can
be written as
\begin{equation}
\eqlabel{PF}
\call = \theta^4 - z (\theta + r_1)(\theta+r_2)(\theta+r_3)
(\theta+r_4)
\end{equation}
where $(r_1,r_2,r_3,r_4)=(\frac 15,\frac 25,\frac 35,\frac 45),
(\frac 16,\frac 26,\frac 46,\frac 56),
(\frac 18,\frac 38,\frac 58,\frac 78),
(\frac 1{10},\frac 3{10},\frac 7{10},\frac 9{10})$ are the ``indices at infinity'',
and $\theta\equiv z \partial_z$.

\subsection{Quintic}

The generic quintic threefold contains $2875$ lines. That number not being divisible 
by $3$, while the Dwork pencil is invariant under cyclic permutation of 
$(x_1,x_2,x_3)$, suggests that there should exist $\zet/3$-invariant lines for generic 
values of $\psi$. It is not very hard to see that there are, up to ${\rm SL}(2,
\complex)$ transformations on homogeneous coordinates $(u,v)$, and conjugacy class 
of cyclic permutation, precisely two different $\zet/3$-equivariant parameterizations 
that are distinguished by whether the determinant of the generator acting on $(u,v)$
is $1$ or a non-trivial cube root of unity, $\omega$. We'll need the former later, while
for the quintic we are left with the general ansatz
\begin{equation}
\eqlabel{later}
x_1 = u + v \,, \quad
x_2 = u + \omega v\,, \quad
x_3 = u + \omega^2 v\,, \quad
x_4 = a\, u \,, \quad
x_5 = b\, u
\end{equation}
where $a,b$ are two parameters that are constrained by the condition that \eqref{later}
be contained in the mirror quintic,
\begin{equation}
ab  \psi = 6\,,\qquad a^5 + b^5 = 27
\end{equation}
The solutions to these equations (and their images under symmetries of the quintic)
yield the van Geemen lines. The original interest of these lines \cite{albanokatz}
was that they allow continuous (unobstructed) deformations for fixed $\psi$.
The global structure of the corresponding families was worked out in \cite{mustatathesis},
see also \cite{cdvv}. One of the results of this analysis is that the only other
lines besides van Geemen's are the coordinate lines, such as
\begin{equation}
\eqlabel{promise}
C_0 = \{ x_1+x_2=0\,,\; x_3+x_4=0\,,\; x_5 =0\}
\end{equation}
It follows from elementary considerations (or the explicit calculations in \cite{mowa}) 
that these coordinate lines have a vanishing inhomogeneity, \ie,
\begin{equation}
\int_{C_0} \beta = 0 
\end{equation}
where $\beta$ is the two-form in \eqref{where}.
Since they are in addition of primitive degree, the $C_0$ (and their integral 
multiples) are ideally suited to serve as reference cycle for the monodromy 
calculations as mentioned in the introduction. All our other examples have 
similar coordinate lines.

On the other hand, the van Geemen lines have a non-trivial Abel-Jacobi image.
This was pointed out via an infinitesimal calculation in \cite{mustatathesis},
while the complete inhomogeneity was determined in \cite{arithmetic} to be
\begin{equation}
\eqlabel{vG}
f_\omega(z) =  \frac{1+2\omega}{(2\pi\ii)^2} \cdot \frac{32}{45}
\cdot \frac{\frac{63}{\psi^5} + \frac{1824}{\psi^{10}} -\frac{512}{\psi^{15}}}
{\bigl(1-\frac{128}{3\psi^5}\bigr)^{5/2}}
\end{equation}
We emphasize that despite appearances, the cycle does not split globally over
$\rationals(\omega)$. When $a=b$ in \eqref{later}, mapping $\omega\mapsto\omega^2$ 
can be compensated by $x_2\leftrightarrow x_3\,,\;x_4\leftrightarrow x_5$, which 
leaves the holomorphic three-form invariant. We see this signaled by the open 
string discriminant $3\psi^5=128$ in the denominator of \eqref{vG}. The other 
branch point is at $\psi=0$.

\medskip

We now briefly review some conics on the mirror quintic found in \cite{arithmetic}.
Consider the $\zet_2\times\zet_2$-invariant ansatz
\begin{multline}
\eqlabel{ansatz}
C_{a,b} = \{ x_1+x_3+ax_5\,,\;\;x_2+x_4+a x_5\,, \\
x_3^2 + x_4^2 + b x_3x_4 + (a+\textstyle{\frac 12} a b) (x_3+x_4)x_5 + 
\textstyle{\frac 18}(-\psi a+6 a^2+2a^2b)
x_5^2\}
\end{multline}
These conics lie on the quintic precisely if
\begin{equation}
\eqlabel{conics}
\begin{split}
64 + 5 a^3 \psi^2 -40 a^4 \psi +12 a^5 &=0 \\
\psi -2 a+a b^2&=0
\end{split}
\end{equation}
These are, for fixed generic $\psi$, 10 different conics, so the covering 
$\hat B\to B$ is quite a bit more interesting than for the van Geemen 
lines. Before discussing it, we note that passing to the global coordinate 
$z=\psi^{-5}$ is easily accomplished since eqs.\ \eqref{conics} are invariant 
under $(\psi,a)\to (\eta \psi,\eta a)$, when $\eta^5=1$. 

Now, the nature of the symmetry $b\to -b$ shows that the 10 conics group as 
pairs of conics in 5 different planes determined by the first equation of 
\eqref{conics}. That symmetry acts trivially when $a=\psi/2$, which under the 
first equation can be seen to coincide with the discriminant locus of the van 
Geemen lines, $3\psi^5=128$. Indeed, at this point, the conics \eqref{ansatz} 
are reducible to two members of the van Geemen family. The conics are also reducible 
at $7\psi^5=128$, but this is not a branch point of the covering \eqref{conics}. 

The discriminant of the first equation in \eqref{conics} is
\begin{equation}
-5308416+26104832 \psi^5+459 \psi^{10} = 0
\end{equation}
In $B$ (parameterized by $z=\psi^{-5}$), these are the two points
\begin{equation}
z_{\pm} = \frac{50986\pm 6875\sqrt{55}}{20736}
\end{equation}
The inhomogeneity corresponding to $C_{a,b}$ was also calculated in \cite{arithmetic}.
The result can be simplified to
\begin{equation}
\eqlabel{fconics}
\begin{split}
&f_{a,b} = \frac{1}{\pi^2}\,\cdot\,
\frac{b}{8640 (2 a-\psi)^5 (12 a^2-32 a \psi+3 \psi^2)^5}\,\cdot \\
& \Bigl(-366917713920-1016582897664 a^4 \psi+3474322882560 a^3 \psi^2
-3601465344000 a^2\psi^3  \\
&+2232487772160 a \psi^4 +1993006776320 \psi^5
-1127509778432 a^4 \psi^6 -62141296640 a^3 \psi^7\\
&+139109736960 a^2 \psi^8
-48377468160 a \psi^9 
+8404041600 \psi^{10}+92770596 a^4 \psi^{11} \\
& -308068920 a^3 \psi^{12}
+34766415 a^2 \psi^{13}+486000 a \psi^{14}\Bigr)
%
%
\end{split}
\end{equation}
As it should be, the three components of the open string discriminant 
\begin{equation}
\eqlabel{shouldbe}
\zDo{11}= z_-\,,\quad 
\zDo{12} = z_+ \,,\quad
\zDo{2} = \frac{3}{128} 
\end{equation}
are manifest in the denominator of the inhomogeneity.

\subsection{Sextic}

This subsection contains the first new results. As is well-known, the generic number of 
lines on a weighted sextic Calabi-Yau threefold is $7884$.\footnote{The mirror formula 
for the number of lines is
$
\frac{d}{\prod w_i}\Bigl[ \frac{d^d}{\prod w_i^{w_i}} -
\frac{d!}{\prod w_i!}\bigl(3 d H_d-3\sum w_i H_{w_i}+ 2\bigr)\Bigr]
$}
Since this is divisible by $3$, it is possible for the cyclic permutations of the 
homogeneous variables, such as $(x_1,x_2,x_3)\to (x_3,x_1,x_2)$ to act freely on
the set of lines. Working out the equations, we find that indeed there are no 
$\zet_3$-invariant lines on (the one-parameter family mirror to) 
$\projective_{11112}^4[6]$. 

As an example that divisibility (of the generic number of solutions by the order of a
symmetry group) does not imply absence of solutions (invariant under that symmetry), 
we consider lines invariant under the $\zet_2$ symmetry
\begin{equation}
\eqlabel{only}
(x_1,x_2,x_3,x_4,x_5)\mapsto (x_2,x_1,x_4,x_3,x_5)
\end{equation}
With a parameterization ansatz
\begin{equation}
\begin{split}
x_1 &= a_1 u + v\,,\qquad x_2 = a_1 u - v \\
x_3 &=u + a_2 v \,,\qquad x_4= u-a_2 v \\
x_5 & = a_3 u^2 + a_4 v^2
\end{split}
\end{equation}
we find the space of such lines factors over $\rationals$ into several components. 
The simplest of those has $a_1$ and $a_2$ equal to sixth roots of $-1$, and $a_3=a_4=
\sqrt{-3\psi}$. These are nothing but the curves studied in \cite{krwa}. The 
corresponding inhomogeneity was found to be proportional to $z^{1/2}$, and the 
monodromy of solutions was also completely worked out. In the present paper, 
we do not wish to discuss these ``toric'' curves any further.

The next more complicated lines with $\zet_2$ symmetry \eqref{only} turn out to also
be invariant under a second $\zet_2$ acting as
\begin{equation}
\eqlabel{z22}
(x_1,x_2,x_3,x_4,x_5)\mapsto (x_3,x_4,x_1,x_2,x_5)
\end{equation}
Imposing \eqref{only} and \eqref{z22} forces the parameters of our ansatz to respect
\begin{equation}
a_2=a_1\,,\qquad a_4=a_3
\end{equation}
Eliminating $a_4$ (and ignoring the toric solutions), the equations boil down to
\begin{equation}
\eqlabel{conjecture}
1-18 a_2^2+111 a_2^4-252 a_2^6+111 a_2^8-18 a_2^{10}+a_2^{12}+\psi^3+15 a_2^4 \psi^3
+32 a_2^6 \psi^3+15 a_2^8\psi^3+a_2^{12} \psi^3 = 0\,,
\end{equation}
and its images under multiplying $a_2$ by a third root of unity.

With these out of the way, we can complete the reduction of the curves with only
a single $\zet_2$ symmetry, \eqref{only}. There are two more components over
$\rationals$. The first is characterized by the vanishing of
\begin{equation}
\eqlabel{vanish}
1+2 a_2^6+a_2^{12}-4 a_2^6 \psi^3
\end{equation}
while the second by
\begin{multline}
\eqlabel{final}
1953125+7812500 a_2^6+11718750 a_2^{12}+7812500 a_2^{18}+1953125 a_2^{24}
+2062500 \psi^3\\ -6250000 a_2^6 \psi^3-16625000 a_2^{12} \psi^3-
6250000 a_2^{18} \psi^3+2062500 a_2^{24} \psi^3+726000 \psi^6\\
-3156000 a_2^6 \psi^6
+12236000 a_2^{12} \psi^6 -3156000 a_2^{18} \psi^6+726000 a_2^{24} \psi^6+85184 \psi^9
\\-484864 a_2^6 \psi^9-180096 a_2^{12} \psi^9-484864 a_2^{18} \psi^9+85184 a_2^{24} \psi^9
+18944 a_2^6 \psi^{12}\\+18688 a_2^{12} \psi^{12}+18944 a_2^{18} \psi^{12}-1024 a_2^{12} \psi^{15}
\end{multline}
We have calculated the inhomogeneity corresponding to \eqref{vanish}, and found it to 
vanish. We suspect the same to hold for \eqref{conjecture}, although we have not
completed the calculation. (The basis for this conjecture is that the discriminant
meets the conifold locus $\psi=1$.)

We have not calculated the inhomogeneity corresponding to \eqref{final}, but presumably it
does not vanish.

Since these results do not yield any new inhomogeneity for the Picard-Fuchs equation of
the sextic, we will drop it from the list for the rest of this paper.

\subsection{Octic}

The number of lines on the weighted octic Calabi-Yau threefold is $29504$. This is
not divisible by $3$, so there should be analogues of the van Geemen lines.
Indeed, let's parameterize lines invariant under
\begin{equation}
\eqlabel{above}
(x_1,x_2,x_3,x_4,x_5)\mapsto (x_2,x_3,x_1,x_4,x_5)
\end{equation}
via
\begin{equation}
\eqlabel{nontrivial}
x_1 = u + v \,,\quad x_2=u+\omega v \,,\quad x_3= u+\omega^2 v\,,\quad
x_4 =a\, u \,,\quad x_5 = b\, u^4+ c\, u v^3
\end{equation}
where $\omega$ is a non-trivial cube root of unity. We find that the space of
such lines factors globally in several components. The first of those has
\begin{equation}
\eqlabel{i1}
a^8=3^4
\end{equation}
while the second
\begin{equation}
\eqlabel{i2}
a^2\psi^2 = 21
\end{equation}
Note that both of these expressions are invariant under $(\psi,a)\to 
(\eta \psi,\eta^{-1}a)$ when $\eta^8=1$, so that the corresponding cycle is 
indeed well-defined over $B$. 

We have calculated the corresponding inhomogeneities, with the following results. 
For the first component, \eqref{i1}, we find:
\begin{equation}
\eqlabel{f1}
f_1(z) = \frac{\sqrt{-3}}{(2\pi\ii)^2}\cdot \frac{3}{16} \cdot
\frac{\psi(8+\psi^2)}{(\psi^2-7)^{5/2}}
\end{equation}
For the convenience of the reader, we explain a few of the intermediate steps leading 
to \eqref{f1} in appendix \ref{few}.

The second component of lines on the octic with the $\zet_3$ symmetry \eqref{above}
gives inhomogeneity:
\begin{equation}
\eqlabel{f2}
f_2(z) = \frac{\sqrt{-7}}{(2\pi\ii)^2}\cdot \frac{147}{16} \cdot
\frac{-823543+184534\psi^8+129\psi^{16}}{\psi^4 (\psi^8-2401)^{5/2}}
\end{equation}

Finally, we note that we have also studied lines with only a $\zet_2$ symmetry
exchanging two coordinates. For all the ones for which we have computed the 
inhomogeneity, it vanishes.

\subsection{Dectic}

Last on the list of one-parameter Calabi-Yau hypersurfaces is the weighted dectic, 
$\projective_{11125}^4[10]$. It contains generically $231200$ lines, a number also
not divisible by $3$. Searching for lines that are invariant under cyclic permutation
of the first three variables, we find that the ansatz analogous to \eqref{later}
allows only $3$ parameters (the coefficient of $u^2$ in $x_4$ and the coefficients of
$u^5$, $u^2v^3$ in $x_5$), constrained by a total of $4$ independent equations
(the coefficients of $u^{10}$, $u^7 v^3$, $u^4 v^6$, $u vß^9$ in $W$), so there are 
generically no solutions. This is where we remember the other possible $\zet_3$-equivariant
parameterization. The ansatz
\begin{equation}
\eqlabel{slightly}
x_1=u+v\,, \quad x_2= \omega u+\omega^2 v\,,\quad x_3 = \omega^2 u+\omega v\,,
\quad x_4= a\, uv\,,\quad x_5 = b\, u^4 v+c\, u v^4
\end{equation}
yields $3$ parameters constrained by $3$ equations (coefficients of
$u^2 v^8$, $u^5v^5$, $u^8 v^2$). We find that there are two components, characterized
by the vanishing of $a^5+3^5$ and $513+a^5-10 a^2 \psi^2$, respectively.
It turns out that the first has vanishing inhomogeneity, while the second gives
\begin{equation}
\eqlabel{segi}
\begin{split}
f(z) &= \frac{1+2\omega}{(2\pi\ii)^2} \cdot  
\frac{9}{50\psi^5(b-a\psi)^5 (a^3-4\psi^2)^5} \cdot \\
& \;\; \Bigl(87483691656-9805676940 a^2 \psi^2 +402856335 a^4 \psi^4
-2843845767 a \psi^6 \\
&\;\;
+448788924 a^3 \psi^8 +10768937688 \psi^{10} 
 -321417648 a^2 \psi^{12}+810896 a^4 \psi^{14}\\
&\;\; -15348960 a \psi^{16}
+299200 a^3 \psi^{18} \Bigr)
\end{split}
\end{equation}
where the parameters $a,b,c,\omega$ in the ansatz \eqref{slightly} are determined
by the system of equations,
\begin{equation}
\eqlabel{ideal}
\begin{split}
513+a^5-10 a^2 \psi^2&=0\,,\\
 27+b^2-2 a b \psi &=0\,,
\end{split}
\end{equation}
as well as $c=2 a \psi-b$, $\omega$: a non-trivial cube root of unity. This system is somewhat
similar to eqs.\ \eqref{conics} for conics on the quintic, in that there are 5
pairs of solutions for fixed $\psi$. The open string discriminant has two components,
\begin{equation}
\zDo{1} =  {\frac{128}{3^6\cdot 19^3}
}\,,\qquad \zDo{2} = {\frac{1}{243}}
\end{equation}
also apparent in \eqref{segi} (recall that in present conventions, $z= \psi^{-10}$).

\section{Algebraic Integrality}
\label{algebraic}

Consider the A-model expansion of the truncated normal function $\tau$ 
\eqref{truncated} associated to an algebraic cycle, around the point of maximal 
unipotent monodromy, $z=\zLV=0$. It is defined as the expansion in the mirror
variable $q$ from eq.\ \eqref{canonical}, of the quantity
\begin{equation}
\calw_A = \frac{\langle \tilde \nu,\cdot\rangle}{\langle\gamma_0,\cdot\rangle}
\Bigl|_{(F^3\calh)^\times}
\end{equation}
On general grounds, explained in the introduction, the A-model expansion takes the 
form
\begin{equation}
\eqlabel{Amodel}
\calw_A = \frac{s}{2\pi\ii r}\,{\log q} + a + \frac{1}{(2\pi\ii)^2}
\sum_{d=1}^\infty \tilde n_d q^{d/r}
\end{equation}
Here, $r$ and $s$ are integers, with $r$ measuring the ramification index of the
cycle at $z=\zLV$ (and $s$ is defined $\bmod r$). Namely, the cycle is really 
defined over the locally extended moduli space with local coordinate $z^{1/r}$. 
Moreover, $a$ is an a priori arbitrary complex constant. 

We emphasize again that the ``classical terms''  in \eqref{Amodel} (the constant 
$a$ and the $\log q$ term) are not determined by the inhomogeneous Picard-Fuchs equation
that we calculated in the previous section. This follows from the definitions
on account of the fact that $\varpi_0$ and $\varpi_1$ (eq.\ \eqref{viz}) are 
periods, \ie, solutions of the homogeneous Picard-Fuchs equation. Instead, the
classical terms 
can be recovered from a monodromy calculation, as we will do in the next section. 
In this section, we will concentrate on the non-trivial part of the $q$-expansion 
in \eqref{Amodel} (the ``instanton expansion'').
This serves two purposes. First, we want to explain the splitting of the
extension $\hat B\to B$ at $\zLV$, and emphasize again that the coefficients 
$\tilde n_d$ are not in general rational numbers. Second, we want to display the 
algebraic integrality discussed in \cite{arithmetic,svw} that is nevertheless inherent 
in the instanton expansion.

\subsection{Octic}

The instanton expansion we find for our lines on the octic is rather similar
to that of the van Geemen lines originally studied in \cite{arithmetic}.
In particular, the field extension is quadratic and appears only as an overall
constant.

\subsubsection*{First Component}

Solving the differential equation with $f_1$ from \eqref{f1} as inhomogeneity,
and doing the expansion, we find, up to the classical terms
\begin{equation}
\eqlabel{upto}
\begin{split}
\frac{(2\pi\ii)^2}{\sqrt{-3}}
\cdot\calw_A &= 768 q^{1/4}+19584 q^{1/2} +860160 q^{3/4}+48733440 q+
\textstyle{\frac{79882616832}{25}} q^{5/4} \\
&+230232655872 q^{3/2}
+
\textstyle{\frac{868448460865536}{49}} q^{7/4}+
1432733965743360 q^2\\
&
+120259506663856128 q^{9/4}+
\textstyle{\frac{259997807371266134016}{25}} q^{5/2} \\
&
+\textstyle{\frac{111494235354933550841856}{121}} q^{11/4}+
83296525620921045651456 q^3+ \cdots
\end{split}
\end{equation}
Note that while the field extension is quadratic, the good local variable is in fact 
$q^{1/4}$ (\ie, $r=4$). Taking account of the sign of the square-root of $-3$, this 
corresponds to 
a degree $8$ covering, locally organized in two groups of degree $4$. This can in fact 
also be seen from the defining eqs.\ \eqref{nontrivial}, \eqref{i1}. Multiplying $a$ 
with an eighth root of unity in general will give a different curve (the calculation 
in the appendix was done for $a^2=3$), but $(a,\omega)\to (-a,\omega^2)$ can be 
compensated by $(x_2,x_3,x_4)\to (x_3,x_2,-x_4)$, an operation that leaves the
holomorphic three-form invariant. 

Since $\sqrt{-3}$ appears only as an overall constant, the expected integrality
takes a fairly simple form. It can be written by twisting the standard Ooguri-Vafa 
multi-cover formula by the corresponding quadratic residue character. Namely, with
\begin{equation}
\eqlabel{asin}
\sum_d \tilde n_d q^{d/4} = \sum_{d,k} n_d \frac{\left(\frac{-3}{k}\right)}{k^2} q^{dk/4}
\end{equation}
where $\left(\frac{-3}{k}\right)=0,1,-1$ if $k=0,1,2\bmod 3$, the $n_d$ are integers (times
$\sqrt{-3}$). For instance,
\begin{equation}
\begin{split}
n_7 &= \tilde n_7 - \textstyle{\frac{\tilde n_1}{49}} = \sqrt{-3}
\cdot 17723437976832 \\
n_{11} &= \tilde n_{11} +\textstyle \frac{\tilde n_1}{121} = 
\sqrt{-3}\cdot 921439961611021081344
\end{split}
\end{equation}

\subsubsection*{Second Component}

This case has $r=1$, and a quadratic extension $\rationals(\sqrt{-7})$ as
residue field. There is no branching at $\zLV$. The first few 
terms of the A-model expansion are
\begin{equation}
\begin{split}
\frac{(2\pi\ii)^2}{\sqrt{-7}} & \calw_A =
77672448 q+2364921695023104 q^2
+139205158983427963682816 q^3 \\
 +1083&3679402194213394854742437888 q^4+ 
\textstyle{\frac{24618206559572019809666493201002121265152}{25}} q^5
+\cdots
\end{split}
\end{equation}
In
\begin{equation}
\sum_d \tilde n_d q^d = \sum_{d,k} n_d \frac{\left(\frac{-7}{k}\right)}{k^2}
q^{k d}
\end{equation}
the $n_d$ are integral,\eg,
\begin{equation}
n_5 = \tilde n_5 +\textstyle{\frac{\tilde n_1}{25}} =
\sqrt{-7}\cdot 984728262382880792386659728040087957504
\end{equation}

\subsection{Dectic}
\label{decticA}

We now turn to the extension \eqref{ideal} of the dectic moduli space, with
corresponding inhomogeneity \eqref{segi}. Because of the high degree of these
equation, we cannot solve them in terms of radicals as we did before. Rather, we
rely on expanding the parameters around $\psi^{-10}=z=\zLV=0$ in a fractional
power series.

Consider first the equation for $a$. The upper boundary of the Newton polygon of 
the first of \eqref{segi} consists of two segments, one of slope $1$, and one of 
slope $-2/3$. The corresponding local field extensions can be read off from the
coefficients on those segments. We see that the power series for $a$ has
coefficients in $\rationals(\sqrt{513/10})$ and $\rationals(10^{1/3})$, respectively. 

We can now insert these solutions into the second of eqs.\ \eqref{segi}. In the
first case, $\rationals(\sqrt{513/10})$ suffers another quadratic extension, 
while in the second, the equation for $b$ splits over $\rationals(10^{1/3})$ 
in the limit $\psi\to\infty$. 

What is not immediately obvious is that, upon plugging these results into the 
inhomogeneity \eqref{segi}, and taking into account the overall prefactor 
$1+2\omega=\sqrt{-3}$, it turns out that the final result for the residue fields 
at $\zLV$ is significantly simpler than at some of the intermediate steps.

Specifically, the first group of four branches has ramification index $r=2$,
and residue field $K_1=\rationals(\zeta_1,\zeta_2)$, where $\zeta_1^2=-2$,
$\zeta_2^2=-57$. The A-model expansion of the normal function is
\begin{equation}
\eqlabel{fgroup}
\begin{split}
(2\pi\ii)^2 \calw_A &=
480120 \zeta_1 q^{1/2}
+2894243400 \zeta_2 q
+\textstyle\frac{3072231093399320}{3} \zeta_1 q^{3/2} \\
+16749&751924576485360 \zeta_2 q^2
+\textstyle\frac{45634140857715370626589476}{5} \zeta_1 q^{5/2} \\
+19269&2509139523826715663010240 \zeta_2 q^3
+\textstyle\frac{6085990469674530883728974279217064500}{49} \zeta_1 q^{7/2} \\
+295&4640183071216785838740930876082745120 \zeta_2 q^4 \\
&+\textstyle\frac{56219253156252289550103460315334373757593346895}{27} \zeta_1 q^{9/2}
+ \cdots
\end{split}
\end{equation} 
Note that in this case, the irrationality of the coefficients is not just an 
overall constant. The twist of the multi-cover formula depends on $n_d$:
\begin{equation}
\eqlabel{around}
\sum_d \tilde n_d q^{d/2} =  \zeta_1 \; \sum_{k,d\;{\rm odd}} 
n_d \frac{\left(\frac{-2}{k}\right)}{k^2}
q^{d k/2}  + \zeta_2 \; \sum_{k,d} n_{2d} \frac{\left(\frac{-57}{k}\right)}{k^2}
q^{d k}
\end{equation}
Moreover, as was already noticed in \cite{arithmetic}, the $n_d$ might not be
integral at the discriminant of the extension $K_1/\rationals$. Here, 
the denominator of $n_d$ for $d$ odd is a growing power of $2$.

For the remaining $6$ branches of our cycle, with $r=3$, the residue 
field in fact collapses back to $\rationals(\sqrt{-3})$. The A-model
expansion is
\begin{equation}
\eqlabel{collapses}
\begin{split}
\frac{(2\pi\ii)^3}{\sqrt{-3}}\calw_A &=
56100 q^{1/3}+35413275 q^{2/3}+42226839000 q
+\textstyle{\frac{264700529287425}{4}} q^{4/3}\\
&+120847065541631256 q^{5/3} 
+243374447043299404350 q^2\\
&+\textstyle{\frac{25706778509839946246266800}{49}} q^{7/3}
+\frac{19022909901384216052391949375}{16} q^{8/3} +\cdots 
\end{split}
\end{equation}
and we have an integrality structure as in \eqref{asin}.

\subsection{Quintic}

So far, all residue field extensions have had abelian Galois group. The simplest
(unfortunately, not simple) example with a non-abelian Galois group that we know
comes from the conics on the mirror quintic. We refer to \cite{arithmetic,svw}
for the full explanation of the integrality, and here content ourselves with
briefly reviewing the branch structure.

As mentioned before, the situation is rather similar to that for lines on the
dectic. The extension by $a$ in \eqref{conics} splits around $\psi\to\infty$ in
one of degree $2$, and one of degree $3$, which are then both extended 
quadratically once we add $b$. In the end, the total residue field extension for 
the first group is bi-quadratic, of the form $\rationals(\zeta)$, with $\zeta^4+
100 \zeta^2-6000=0$. The Galois group of this polynomial is the (non-abelian)
dihedral group $D_4$. Note in particular that this extension survives in the
expansion of the inhomogeneity \eqref{fconics}. The A-model expansion is
\begin{equation}
\eqlabel{survives}
\begin{split}
(2\pi\ii)^2\cdot \calw_A &= 
\textstyle
 (-304960000 \zeta+7227200 \zeta^3) q \\
\textstyle
-\frac{512000}{51} & (-1016270788225 \zeta+24084846092 \zeta^3) q^2 \\
\textstyle
+\frac{40000}{7803}  (-13&1215286737935072263800 \zeta+3109702672077500263451 \zeta^3)q^3
+ \cdots
\end{split}
\end{equation}
The second group of conics around $\zLV$ has $r=6$, and residue 
field $\rationals (5^{1/3})$.
\begin{equation}
\eqlabel{secondgroup}
\begin{split}
(2\pi\ii)^2\cdot \calw_A &= 
\textstyle 2400 \cdot 5^{2/3}  q^{1/6}-400600 q^{1/2}
+\frac{120620000}{3}\cdot 5^{1/3} q^{5/6}
\\
&\qquad\qquad\textstyle -\frac{7863785008000}{1323}\cdot 5^{2/3} q^{7/6}
+\frac{48067627724000}{9} q^{3/2}+\cdots
\end{split}
\end{equation}

\section{Integrality of Monodromy}
\label{topological}

As before, we first collect the homogeneous data, with implicit reference to the
introduction of the paper.

The point of the normalization \eqref{convenient} of the holomorphic three-form is 
that the so-called fundamental period \cite{cdgp} takes a particularly compact 
form---the three-cycle $\gamma_0$, defined in a neighborhood of $z=\zLV=0$ 
by encircling the coordinate axes, is invariant under $z\to \ee^{2\pi\ii}z$ and
gives the period
\begin{equation}
\varpi_0(z) = \int_{\gamma_0} \Omega = \frac{1}{(2\pi\ii)^5}\int \prod_i \frac{dx_i}{x_i}
\sum_{n=0}^\infty \biggl(\frac{\sum \frac{w_i}{d} x_i^{d/w_i}}{\psi{\prod x_i}} \biggr)^n 
= \sum_{n=0}^\infty \frac{(dn)!}{\prod (w_i n)!} \tilde{z}^n
\end{equation}
where $\tilde z = \frac{\prod_i w_i^{w_i}}{d^d} z$. This
can be readily verified to satisfy the Picard-Fuchs equation, \eqref{PF}.
All solutions of that equation around $z=0$ are obtained from the 
hypergeometric generating function,
\begin{equation}
\varpi(\tilde z;H) = \sum_{n=0}^\infty 
\frac{\Gamma\bigl(1+ d (n+H)\bigr)}{\prod \Gamma\bigl(1+w_i(n+H)\bigr)}
\tilde{z}^{n+H}
\end{equation}
by taking derivatives with respect to $H$. We define for $j=0,1,2,3$
\begin{equation}
\varphi_j(z) = \frac{1}{(2\pi\ii)^j} (\partial_H)^j\bigr|_{H=0} \varpi(z;H)
\end{equation}
In terms of these, the integral basis of periods is given by%
\footnote{This basis is almost equal to $\frac{1}{j!(2\pi\ii)^j}(\del_H)^j|_{H=0}
\frac{\prod\Gamma(1+w_i H)}{\Gamma(1+d H)}\varpi(z;H)$, with due account of
$\kappa$. Integral monodromy still prefers \eqref{interms}.}
\begin{equation}
\eqlabel{interms}
\Pi = \begin{pmatrix} 
\varpi_0\\ \varpi_1 \\ \varpi_2 \\ \varpi_3\end{pmatrix}
= \begin{pmatrix} 
1 & 0 & 0 & 0 \\
0 & 1 & 0 & 0 \\
-\frac {c_2}{24} & -\frac{\kappa}{2} & \frac{\kappa}{2} & 0 \\
\alpha & -\frac{c_2}{24} & 0  & -\frac{\kappa}{6} \end{pmatrix}
\cdot
\begin{pmatrix}
\varphi_0 \\ \varphi_1 \\ \varphi_2 \\ \varphi_3 
\end{pmatrix}
\end{equation}
Here, the change of basis involves some topological invariants of the mirror manifold,
\viz, the classical triple intersection number $\kappa=d/\prod w_i$, the second
Chern number $c_2= \kappa \sum w_iw_j$ and the Euler number $\chi=\kappa
(\sum w_iw_jw_k - d\sum w_iw_j)$, which enters the constant \eqref{ingeneral},
\footnote{For $d=5,6,8,10$, $\kappa=5,3,2,1$, $c_2=50,42,44,34$, $-\chi=200,204,296,288$.}
\begin{equation}
\alpha = -\chi \frac{\zeta(3)}{(2\pi\ii)^3}
\end{equation}
With respect to this basis, the large volume and conifold monodromy are given
by the matrices
\begin{equation}
\MLV = 
\begin{pmatrix} 
1 & 0 & 0 & 0 \\ 1 & 1 & 0 & 0 \\ 0 & \kappa & 1 & 0 \\
-\frac{c_2}{12}-\frac\kappa6 & -\kappa & -1 & 1 
\end{pmatrix}
\qquad
\MC =
\begin{pmatrix}
1 & 0 & 0 & 1 \\ 0 & 1 & 0 & 0 \\ 0 & 0 & 1 & 0 \\ 0 & 0 & 0 & 1 
\end{pmatrix}
\end{equation}
while the Gepner monodromy is
\begin{equation}
\eqlabel{gepner}
\MG = \MLV\cdot \MC
\end{equation}
The general strategy for calculating the extension of these matrices by the
algebraic cycles is explained at the beginning of subsection \ref{dectic}.
First we warm up with those somewhat simpler examples.

\subsection{Octic}

As we have seen in the previous sections, the extended moduli spaces associated
with lines on the octic are rather similar to that of the van Geemen lines on 
the quintic, studied in \cite{lawa}. 

\subsubsection*{First component}

More precisely, the lines with inhomogeneity \eqref{f1} are associated with a
degree 8 covering, which is branched at $\zLV$ with ramification index $4$
and at the open string discriminant $\zD=7^{-4}$ with index $2$. In other words, 
at $\zLV$, the 8 branches split into two groups of four, while at $\zD$, the $8$ 
branches come together pairwise.

To give a little more detailed account of what is going on globally, we 
recall that the inhomogeneity \eqref{f1} was calculated (see appendix) 
over $\psi$-space as corresponding to the solution $a^2=3$ of \eqref{i1}. Because 
of the choice of root of $-3$, these are really two branches of our cycle. The other 
branches can be reached by multiplying $a^2$ by a fourth root of unity,
or equivalently by a monodromy $z\to \ee^{2\pi\ii}z$.

Now notice that all interesting points in $z$-space line up conveniently on
the real axis: $0=\zLV<\zD<\zC < \zG=\infty$. Therefore, it is natural to
carry out the monodromy calculations along the positive real axis, where
we write $f_1$ as
\begin{equation}
\eqlabel{their}
\frac{\sqrt{-3}}{(2\pi\ii)^2}
\cdot \frac{3}{16} \cdot \frac{z^{1/4} + 8 z^{1/2}}{(1-7 z^{1/4})^{5/2}}
\end{equation} 
Let's label the lines with this calculated inhomogeneity on the positive
real axis as $C_{z,1}$ and  $C_{z,2}$ for the two choices of square-root
respectively, and the associated truncated normal functions $\tau_1$, $\tau_2$. 
Recall that it is understood implicitly that we are calculating the chain 
integrals with respect to some fixed globally constant cycle (\eg, a 
``coordinate line''). Naturally, the other lines and normal functions 
would be labeled $C_{z,k}$, $\tau_k$ with $k=3,\ldots 8$, and their
inhomogeneity is obtained from \eqref{their} by multiplying $z^{1/4}$
with a fourth root of unity.

In accord with this geometric situation, we can fix the ambiguity in the
solution of the inhomogeneous equation, and hence recover the full $\tau_k$'s, 
by imposing that for $k=1,2,3,4$, we have $\tau_{2k-1} = -\tau_{2k}$, and that 
they vanish at $z=\zD$. This is the same strategy as in \cite{lawa}.
In the local coordinate $y=1-7 z^{1/4}$, we find
\begin{equation}
\eqlabel{both}
\frac{4\pi^2}{\sqrt{-3}}\tau_1 = 
\frac{392}{15} y^{3/2}+\frac{135191}{5625} y^{5/2}
+\frac{23856287}{1125000} y^{7/2} +\cdots
\end{equation}
To determine the behaviour of $\tau_1$ at $\zLV$, we might continue it numerically
as a solution of the differential equation, or as in \cite{lawa}, pick 
a convenient point of comparison between $\zLV$ and $\zD$ where both
the power series expansion \eqref{both} and
\begin{equation}
\frac{(2\pi\ii)^2}{\sqrt{-3}} \tau_1^{\rm LV} =
48 z^{1/4}+\frac{153}{2}z^{1/2}+210 z^{3/4}+\frac{190365}{256} z + \cdots
\end{equation}
converge well. We find
\begin{equation}
\tau_1 = \tau_1^{\rm LV} + a \varpi_0 -2\varpi_1 + \textstyle\frac{1}{2} \varpi_3
\end{equation}
where  
  $a\approx \ii 3.085052546678470732727\ldots $

\subsubsection*{Second component}

The analysis for $f_2$ is quite similar, the main difference being that the covering
has total degree $2$ and is branched only at $\zD$ and $\zG$. The expansions
are
\begin{equation}
\frac{(2\pi\ii)^2}{\sqrt{-7}} \tau_1
=
\frac 75 y^{3/2}+
\frac{23051}{20000} y^{5/2}
+\frac{15388807}{16000000} y^{7/2}
+\frac{19044150391}{23040000000} y^{9/2} + \cdots
\end{equation}
(where $y=1-2401 z$) and
\begin{equation}
\frac{(2\pi\ii)^2}{\sqrt{-7}} \tau_1^{\rm LV} = 
\frac{18963}{16} z + \frac{577705085643}{1048576}z^2+
\frac{2125304249123593811}{4294967296} z^3
+ \cdots
\end{equation}
Analytic continuation shows
\begin{equation}
\tau_1 = \tau_1^{\rm LV} + a \varpi_0 - 3 \varpi_1 + \textstyle\frac 12 \varpi_3
\end{equation}
with 
$a\approx \ii 6.48474571034689069\ldots $

\subsection{Dectic}
\label{dectic}

We are now ready to embark on the calculation of the monodromy for the inhomogeneity 
\eqref{segi}. As discussed previously, the covering has total degree $10$, branched at 
$\zLV$, $\zDo{1}$, and $\zDo{2} $. Again, all these branch points lie on the real axis, 
with
\begin{equation}
\eqlabel{lineup}
0 = \zLV < \zDo1 (\approx 2.56\cdot 10^{-5}) < 
\zDo2 (\approx 4.12\cdot 10^{-3})
<\zC = 1 < \zG
\end{equation}
Recall the basic goal and strategy: We want to determine the asymptotic behaviour 
at $\zLV$ of the truncated normal function $\tau_k$ for $k=1,\ldots 10$, associated 
with each branch, $C_{z,k}$ of our algebraic cycle. To this end, we need to fix the 
solution of the inhomogeneous Picard-Fuchs equation modulo integral periods. Pertinent
information is contained in the boundary condition at the open string discriminant,
and in the statement that all monodromies be integral. The simplest degeneration 
(which is all we have to deal with in our examples) is that two branches, say 
$C_{z,k_1}$ and $C_{z,k_2}$, come together at a component of the discriminant, say
$\zD$. Then the condition is that there be an integral period $p$ such that
\begin{equation}
\eqlabel{vanishing}
\frac{\tau_{k_1} - \tau_{k_2} - p}{(z-\zD)^{3/2}}
\end{equation}
is regular at $\zD$. As explained in \cite{lawa}, this condition ensures that the 
full normal function vanishes at $\zD$ (in other words, not only the integral of
the holomorphic three-form over a bounding three-chain, but also its derivative,
which gives the integral of the $(2,1)$-form). Under monodromy around $\zD$,
we have
\begin{equation}
\eqlabel{saw}
(\tau_{k_1},\tau_{k_2}) \to (\tau_{k_2}+p,\tau_{k_1}-p)
\end{equation}
We will refer to \eqref{vanishing} as the ``vanishing domain wall'' condition.
As we just saw, it ensures integrality of monodromy around the open string
discriminant. 

In general, integral monodromy is the statement that for each of our singular
points (including open string discriminant, $\zLV$, $\zC$ and $\zG$), there should be 
a permutation matrix $(\sigma_k^l)$ and an integral matrix $(A_k^i)$ such that upon 
encircling that point,
\begin{equation}
\eqlabel{fullcon}
\tau_k \to \sigma_k^l \tau_{l} + A_{k}^i \varpi_i
\end{equation}
where $(\varpi_i)_{i=0,1,2,3}$ is the integral basis of periods \eqref{interms}. We 
find it convenient to combine $(\sigma_k^l)$ and $(A_k^i)$ with the matrix $M$ 
representing the monodromy of periods into a single matrix $\hat M$ of block form
\begin{equation}
\hat M = \begin{pmatrix} \sigma & A \\ 0 & M \end{pmatrix}
\end{equation}
that acts on the ``extended period vector''
\begin{equation}
\hat \Pi = (\tau_1,\ldots, \tau_{10},\varpi_0,\ldots,\varpi_3)^T
\end{equation}
The collection of these matrices over all singular points gives the extended monodromy 
representation advertised in \eqref{underlies}.

In the previous examples, we exploited the fact that the open string discriminant 
consisted of only a single point, and that there was, up to simple symmetries, 
essentially only one vanishing domainwall. This allowed us to fix $\tau_k$ at the
open string discriminant, and then continue it to $\zLV$ in order to extract
the asymptotic behaviour.

In the present example, we will proceed the other way around. We begin with 
introducing the ``large volume solutions'', $\tau^{\rm LV}_k$, distinguished
by their vanishing at $z=0$. (Technically, we impose that there be no logarithmic
terms, and no constant, in the solution of the differential equation.) We then
calculate the monodromy of these solutions around all the singular points. This
will not be integral in general, but we can improve on this by adding suitable
combinations of the integral periods. Namely, there is a matrix $(B_k^i)$,
which as it turns out is unique modulo integers, such that
\begin{equation}
\eqlabel{unique}
\tau_k = \tau_k^{\rm LV} + B_k^i \varpi_i
\end{equation}
has integral monodromy. At the end, we check all vanishing domainwall conditions
\eqref{vanishing}.

Let us see what this looks like in practice. Referring to subsection \ref{decticA},
we label the four branches in the first group, see eq.\ \eqref{fgroup}, such that 
if $\tau_1^{\rm LV}$ corresponds to the roots $(\zeta_1,\zeta_2)$ of $\zeta_1^2=-2$, 
$\zeta_2^2=-57$, then $\tau_2^{\rm LV}$ corresponds to $(-\zeta_1,\zeta_2)$, 
$\tau_3^{\rm LV}$ to $(-\zeta_1,-\zeta_2)$, and $\tau_4^{\rm LV}$ to 
$(\zeta_1,-\zeta_2)$. Thus, large volume monodromy acts by exchanging 
$(\tau_1^{\rm LV},\tau_2^{\rm LV})$ and $(\tau_3^{\rm LV},\tau_4^{\rm LV})$.

In the second group, see eq.\ \eqref{collapses}, $\tau_{5,6,7}^{\rm LV}$ correspond 
to one choice of $\sqrt{-3}$, and $\tau_{8,9,10}^{\rm LV}$ to the other, ordered 
in the same way such that large volume monodromy acts by cyclic permutation.

Tracking these solutions to the first component of the discriminant $\zDo{1}$
along the positive real axis, we find that it is the combinations 
$(k_1,k_2)=(4,7)$ and $(k_1,k_2)=(1,10)$ that should vanish there (though they
do not quite yet).

Skirting around $\zDo{1}$ in the positive upper half plane, we proceed to
$\zDo{2}$ and find that the vanishing domainwall there will come from the 
combination $(k_1,k_2)=(2,3)$.

Finally, we head for the conifold, encircle it in the positive direction 
and return to $\zLV$ along the
same path. The net result is that the $\tau^{\rm LV}_k$ pick up a complex multiple
of the fundamental period $\varpi_0$, in the above order given by 
\begin{equation}
\textstyle
\Bigl(\frac{9}{2}+ a_1, -3 +a_1, 3- a_1, -\frac 92 -a_1,
-\frac {11}{6} + a_2,
\frac{11}{6}+a_2,
\frac{9}{2} + a_2 ,
\frac{11}{6} - a_2,
-\frac{11}{6} - a_2,
-\frac{9}{2} - a_2\Bigr)
\end{equation}
where $a_1\approx \ii 5.154774632407\ldots$, and 
$a_2\approx \ii 5.090336702019\ldots $.

It is then not hard to check that the change of basis \eqref{unique}, with $B$ given by
\begin{equation}
B^T=
\arraycolsep=2pt
\left(
\begin{array}{cccccccccc}
 -\frac{1}{2}\!-\!a_1 & -a_1 & a_1 & -\frac{1}{2}\!+\!a_1 & -\frac{1}{6}\!-\!a_2  & 
\frac{1}{6}\!-\!a_2  & \frac{1}{2}\!-\!a_2  & \frac{1}{6}\!+\!a_2  & -\frac{1}{6}\!+\!a_2  
& -\frac{1}{2}\!+\!a_2  \\ 
 \frac{1}{2} & \frac{1}{2} & \frac{1}{2} & \frac{1}{2} & \frac{2}{3} & 
\frac{2}{3} & \frac{2}{3} & \frac{1}{3} & \frac{1}{3} & \frac{1}{3} \\ 
 0 & 0 & 0 & 0 & 0 & 0 & 0 & 0 & 0 & 0 \\
 0 & 0 & 0 & 0 & 0 & 0 & 0 & 0 & 0 & 0 \\
\end{array}
\right)
\end{equation}
makes the conifold monodromy integral in a way that is consistent with the 
extension at large volume. Specifically, in this basis the extended monodromy 
matrices are
\begin{equation}
\hat{M}_{\rm LV} =
\mbox{\tiny\ensuremath {\arraycolsep=2pt
\left(
\begin{array}{cccccccccccccc}
 0 & 1 & 0 & 0 & 0 & 0 & 0 & 0 & 0 & 0 & 0 & 0 & 0 & 0 \\
 1 & 0 & 0 & 0 & 0 & 0 & 0 & 0 & 0 & 0 & 1 & 0 & 0 & 0 \\
 0 & 0 & 0 & 1 & 0 & 0 & 0 & 0 & 0 & 0 & 1 & 0 & 0 & 0 \\
 0 & 0 & 1 & 0 & 0 & 0 & 0 & 0 & 0 & 0 & 0 & 0 & 0 & 0 \\
 0 & 0 & 0 & 0 & 0 & 0 & 1 & 0 & 0 & 0 & 0 & 0 & 0 & 0 \\
 0 & 0 & 0 & 0 & 1 & 0 & 0 & 0 & 0 & 0 & 1 & 0 & 0 & 0 \\
 0 & 0 & 0 & 0 & 0 & 1 & 0 & 0 & 0 & 0 & 1 & 0 & 0 & 0 \\
 0 & 0 & 0 & 0 & 0 & 0 & 0 & 0 & 0 & 1 & 1 & 0 & 0 & 0 \\
 0 & 0 & 0 & 0 & 0 & 0 & 0 & 1 & 0 & 0 & 0 & 0 & 0 & 0 \\
 0 & 0 & 0 & 0 & 0 & 0 & 0 & 0 & 1 & 0 & 0 & 0 & 0 & 0 \\
 0 & 0 & 0 & 0 & 0 & 0 & 0 & 0 & 0 & 0 & 1 & 0 & 0 & 0 \\
 0 & 0 & 0 & 0 & 0 & 0 & 0 & 0 & 0 & 0 & 1 & 1 & 0 & 0 \\
 0 & 0 & 0 & 0 & 0 & 0 & 0 & 0 & 0 & 0 & 0 & 1 & 1 & 0 \\
 0 & 0 & 0 & 0 & 0 & 0 & 0 & 0 & 0 & 0 & -3 & -1 & -1 & 1
\end{array}
\right)}}
\qquad
\hat{M}_{{\rm C}}=
\mbox{\tiny\ensuremath{\arraycolsep=2pt
\left(
\begin{array}{cccccccccccccc}
 1 & 0 & 0 & 0 & 0 & 0 & 0 & 0 & 0 & 0 & 0 & 0 & 0 & 4 \\
 0 & 1 & 0 & 0 & 0 & 0 & 0 & 0 & 0 & 0 & 0 & 0 & 0 & -3 \\
 0 & 0 & 1 & 0 & 0 & 0 & 0 & 0 & 0 & 0 & 0 & 0 & 0 & 3 \\
 0 & 0 & 0 & 1 & 0 & 0 & 0 & 0 & 0 & 0 & 0 & 0 & 0 & -5 \\
 0 & 0 & 0 & 0 & 1 & 0 & 0 & 0 & 0 & 0 & 0 & 0 & 0 & -2 \\
 0 & 0 & 0 & 0 & 0 & 1 & 0 & 0 & 0 & 0 & 0 & 0 & 0 & 2 \\
 0 & 0 & 0 & 0 & 0 & 0 & 1 & 0 & 0 & 0 & 0 & 0 & 0 & 5 \\
 0 & 0 & 0 & 0 & 0 & 0 & 0 & 1 & 0 & 0 & 0 & 0 & 0 & 2 \\
 0 & 0 & 0 & 0 & 0 & 0 & 0 & 0 & 1 & 0 & 0 & 0 & 0 & -2 \\
 0 & 0 & 0 & 0 & 0 & 0 & 0 & 0 & 0 & 1 & 0 & 0 & 0 & -5 \\
 0 & 0 & 0 & 0 & 0 & 0 & 0 & 0 & 0 & 0 & 1 & 0 & 0 & 1 \\
 0 & 0 & 0 & 0 & 0 & 0 & 0 & 0 & 0 & 0 & 0 & 1 & 0 & 0 \\
 0 & 0 & 0 & 0 & 0 & 0 & 0 & 0 & 0 & 0 & 0 & 0 & 1 & 0 \\
 0 & 0 & 0 & 0 & 0 & 0 & 0 & 0 & 0 & 0 & 0 & 0 & 0 & 1
\end{array}
\right)}}
\end{equation}
\begin{equation}
\hat{M}_{{\rm D}_1}= 
\mbox{\tiny\ensuremath{\arraycolsep=2pt
\left(
\begin{array}{cccccccccccccc}
 0 & 0 & 0 & 0 & 0 & 0 & 0 & 0 & 0 & 1 & 0 & -4 & 0 & 1 \\
 0 & 1 & 0 & 0 & 0 & 0 & 0 & 0 & 0 & 0 & 0 & 0 & 0 & 0 \\
 0 & 0 & 1 & 0 & 0 & 0 & 0 & 0 & 0 & 0 & 0 & 0 & 0 & 0 \\
 0 & 0 & 0 & 0 & 0 & 0 & 1 & 0 & 0 & 0 & -1 & 4 & 0 & -1 \\
 0 & 0 & 0 & 0 & 1 & 0 & 0 & 0 & 0 & 0 & 0 & 0 & 0 & 0 \\
 0 & 0 & 0 & 0 & 0 & 1 & 0 & 0 & 0 & 0 & 0 & 0 & 0 & 0 \\
 0 & 0 & 0 & 1 & 0 & 0 & 0 & 0 & 0 & 0 & 1 & -4 & 0 & 1 \\
 0 & 0 & 0 & 0 & 0 & 0 & 0 & 1 & 0 & 0 & 0 & 0 & 0 & 0 \\
 0 & 0 & 0 & 0 & 0 & 0 & 0 & 0 & 1 & 0 & 0 & 0 & 0 & 0 \\
 1 & 0 & 0 & 0 & 0 & 0 & 0 & 0 & 0 & 0 & 0 & 4 & 0 & -1 \\
 0 & 0 & 0 & 0 & 0 & 0 & 0 & 0 & 0 & 0 & 1 & 0 & 0 & 0 \\
 0 & 0 & 0 & 0 & 0 & 0 & 0 & 0 & 0 & 0 & 0 & 1 & 0 & 0 \\
 0 & 0 & 0 & 0 & 0 & 0 & 0 & 0 & 0 & 0 & 0 & 0 & 1 & 0 \\
 0 & 0 & 0 & 0 & 0 & 0 & 0 & 0 & 0 & 0 & 0 & 0 & 0 & 1
\end{array}
\right)}}
\qquad
\hat{M}_{{\rm D}_2}=
\mbox{\tiny\ensuremath{\arraycolsep=2pt
\left(
\begin{array}{cccccccccccccc}
 1 & 0 & 0 & 0 & 0 & 0 & 0 & 0 & 0 & 0 & 0 & 0 & 0 & 0 \\
 0 & 0 & 1 & 0 & 0 & 0 & 0 & 0 & 0 & 0 & 0 & -3 & 0 & 0 \\
 0 & 1 & 0 & 0 & 0 & 0 & 0 & 0 & 0 & 0 & 0 & 3 & 0 & 0 \\
 0 & 0 & 0 & 1 & 0 & 0 & 0 & 0 & 0 & 0 & 0 & 0 & 0 & 0 \\
 0 & 0 & 0 & 0 & 1 & 0 & 0 & 0 & 0 & 0 & 0 & 0 & 0 & 0 \\
 0 & 0 & 0 & 0 & 0 & 1 & 0 & 0 & 0 & 0 & 0 & 0 & 0 & 0 \\
 0 & 0 & 0 & 0 & 0 & 0 & 1 & 0 & 0 & 0 & 0 & 0 & 0 & 0 \\
 0 & 0 & 0 & 0 & 0 & 0 & 0 & 1 & 0 & 0 & 0 & 0 & 0 & 0 \\
 0 & 0 & 0 & 0 & 0 & 0 & 0 & 0 & 1 & 0 & 0 & 0 & 0 & 0 \\
 0 & 0 & 0 & 0 & 0 & 0 & 0 & 0 & 0 & 1 & 0 & 0 & 0 & 0 \\
 0 & 0 & 0 & 0 & 0 & 0 & 0 & 0 & 0 & 0 & 1 & 0 & 0 & 0 \\
 0 & 0 & 0 & 0 & 0 & 0 & 0 & 0 & 0 & 0 & 0 & 1 & 0 & 0 \\
 0 & 0 & 0 & 0 & 0 & 0 & 0 & 0 & 0 & 0 & 0 & 0 & 1 & 0 \\
 0 & 0 & 0 & 0 & 0 & 0 & 0 & 0 & 0 & 0 & 0 & 0 & 0 & 1
\end{array}
\right)}}
\end{equation}
One may then first of all make the consistency check that the extension of \eqref{gepner},
\begin{equation}
\hat{M}_{\rm G} = \hat{M}_{\rm LV}\cdot \hat{M}_{{\rm D}_1}\cdot
\hat{M}_{{\rm D}_2} \cdot \hat{M}_{\rm C}
\end{equation}
(remember the lineup \eqref{lineup} and that these matrices compose on the right)
satisfies
$$
(\hat{M}_{\rm G})^{10} = 1
$$
Finally, we verify the existence of vanishing domain walls. We find that
\begin{equation}
\tau_7-\tau_4 -\varpi_0+4\varpi_1-\varpi_3
\qquad\text{and}\qquad
\tau_{10}- \tau_1 -4\varpi_1+\varpi_3
\end{equation}
vanish at $\zDo{1}$, and
\begin{equation}
\tau_2 - \tau_3 + 3\varpi_1
\end{equation}
vanishes at $\zDo{2}$. So everything appears in order.


\subsection{Quintic}

We now repeat those calculations for the extension \eqref{fconics} of
the quintic. The first thing to note is that while the components of the open string
discriminant \eqref{shouldbe} are still all on the real axis, one of them is negative.
Namely
\begin{equation}
\eqlabel{namely}
\zDo{11} (\approx -1.76\cdot 10^{-5})
< \zLV < 
\zDo{2} (\approx 2.34\cdot 10^{-2})
< \zC < 
\zDo{12} (\approx 4.92)
\end{equation}
We label the four branches in the first group eq.\ \eqref{survives} such
that $\tau_1$, $\tau_4$ correspond to the real roots of 
$\zeta^4+100\zeta^2-6000$, and $\tau_2$, $\tau_3$ to the imaginary
roots. Large volume monodromy leaves these branches untouched. Those in the second group, 
\eqref{secondgroup}, are arranged by Mathematica such that large volume monodromy acts by 
$(5,6,7,8,9,10)\to (7,5,9,6,10,8)$. Under conifold monodromy, these solutions 
pick up the fundamental period times
\begin{equation}
\Bigl(a_1,24-a_2,-24+a_2,-a_1,-\frac{59}{2},
-\frac{69}{2}, -\frac{69}{2}, \frac{69}{2} ,
\frac{69}{2} ,\frac{59}{2}\Bigr)
\end{equation}
with $a_1\approx 1.6377482972\ldots$, $a_2\approx
\ii 93.620780658\ldots$. To compensate for this, we add the periods
\begin{equation}
B^T=
\arraycolsep=3pt
\left(
\begin{array}{cccccccccc}
- a_1 & -24+a_2 & 24-a_2 & a_1 & \frac{59}{2} & \frac{69}{2} 
& \frac{69}{2} & -\frac{69}{2} & -\frac{69}{2} & -\frac{59}{2} \\
 0 & 0 & 0 & 0 & 0 & 0 & 0 & 0 & 0 & 0 \\
 0 & 0 & 0 & 0 & 0 & 0 & 0 & 0 & 0 & 0 \\
 0 & 0 & 0 & 0 & 0 & 0 & 0 & 0 & 0 & 0
\end{array}
\right)
\end{equation}
giving the monodromy matrices
\begin{equation}
\hat{M}_{\rm LV} =
\mbox{\tiny\ensuremath {\arraycolsep=2pt
\left(
\begin{array}{cccccccccccccc}
 1 & 0 & 0 & 0 & 0 & 0 & 0 & 0 & 0 & 0 & 0 & 0 & 0 & 0 \\
 0 & 1 & 0 & 0 & 0 & 0 & 0 & 0 & 0 & 0 & 0 & 0 & 0 & 0 \\
 0 & 0 & 1 & 0 & 0 & 0 & 0 & 0 & 0 & 0 & 0 & 0 & 0 & 0 \\
 0 & 0 & 0 & 1 & 0 & 0 & 0 & 0 & 0 & 0 & 0 & 0 & 0 & 0 \\
 0 & 0 & 0 & 0 & 0 & 0 & 1 & 0 & 0 & 0 & -5 & 0 & 0 & 0 \\
 0 & 0 & 0 & 0 & 1 & 0 & 0 & 0 & 0 & 0 & 5 & 0 & 0 & 0 \\
 0 & 0 & 0 & 0 & 0 & 0 & 0 & 0 & 1 & 0 & 69 & 0 & 0 & 0 \\
 0 & 0 & 0 & 0 & 0 & 1 & 0 & 0 & 0 & 0 & -69 & 0 & 0 & 0 \\
 0 & 0 & 0 & 0 & 0 & 0 & 0 & 0 & 0 & 1 & -5 & 0 & 0 & 0 \\
 0 & 0 & 0 & 0 & 0 & 0 & 0 & 1 & 0 & 0 & 5 & 0 & 0 & 0 \\
 0 & 0 & 0 & 0 & 0 & 0 & 0 & 0 & 0 & 0 & 1 & 0 & 0 & 0 \\
 0 & 0 & 0 & 0 & 0 & 0 & 0 & 0 & 0 & 0 & 1 & 1 & 0 & 0 \\
 0 & 0 & 0 & 0 & 0 & 0 & 0 & 0 & 0 & 0 & 0 & 5 & 1 & 0 \\
 0 & 0 & 0 & 0 & 0 & 0 & 0 & 0 & 0 & 0 & -5 & -5 & -1 & 1
\end{array}
\right)}}
\qquad
\hat{M}_{{\rm C}}=
\mbox{\tiny\ensuremath{\arraycolsep=2pt
\left(
\begin{array}{cccccccccccccc}
 1 & 0 & 0 & 0 & 0 & 0 & 0 & 0 & 0 & 0 & 0 & 0 & 0 & 0 \\
 0 & 1 & 0 & 0 & 0 & 0 & 0 & 0 & 0 & 0 & 0 & 0 & 0 & 0 \\
 0 & 0 & 1 & 0 & 0 & 0 & 0 & 0 & 0 & 0 & 0 & 0 & 0 & 0 \\
 0 & 0 & 0 & 1 & 0 & 0 & 0 & 0 & 0 & 0 & 0 & 0 & 0 & 0 \\
 0 & 0 & 0 & 0 & 1 & 0 & 0 & 0 & 0 & 0 & 0 & 0 & 0 & 0 \\
 0 & 0 & 0 & 0 & 0 & 1 & 0 & 0 & 0 & 0 & 0 & 0 & 0 & 0 \\
 0 & 0 & 0 & 0 & 0 & 0 & 1 & 0 & 0 & 0 & 0 & 0 & 0 & 0 \\
 0 & 0 & 0 & 0 & 0 & 0 & 0 & 1 & 0 & 0 & 0 & 0 & 0 & 0 \\
 0 & 0 & 0 & 0 & 0 & 0 & 0 & 0 & 1 & 0 & 0 & 0 & 0 & 0 \\
 0 & 0 & 0 & 0 & 0 & 0 & 0 & 0 & 0 & 1 & 0 & 0 & 0 & 0 \\
 0 & 0 & 0 & 0 & 0 & 0 & 0 & 0 & 0 & 0 & 1 & 0 & 0 & 1 \\
 0 & 0 & 0 & 0 & 0 & 0 & 0 & 0 & 0 & 0 & 0 & 1 & 0 & 0 \\
 0 & 0 & 0 & 0 & 0 & 0 & 0 & 0 & 0 & 0 & 0 & 0 & 1 & 0 \\
 0 & 0 & 0 & 0 & 0 & 0 & 0 & 0 & 0 & 0 & 0 & 0 & 0 & 1
\end{array}
\right)}}
\end{equation}
\begin{equation}
\hat{M}_{{\rm D_{11}}}=
\mbox{\tiny\ensuremath{\arraycolsep=2pt
\left( 
\begin{array}{cccccccccccccc}
 1 & 0 & 0 & 0 & 0 & 0 & 0 & 0 & 0 & 0 & 0 & 0 & 0 & 0 \\
 0 & 0 & 0 & 0 & 0 & 0 & 1 & 0 & 0 & 0 & -96 & 55 & -2 & -4 \\
 0 & 0 & 0 & 0 & 0 & 0 & 0 & 1 & 0 & 0 & 96 & -55 & 2 & 4 \\
 0 & 0 & 0 & 1 & 0 & 0 & 0 & 0 & 0 & 0 & 0 & 0 & 0 & 0 \\
 0 & 0 & 0 & 0 & 1 & 0 & 0 & 0 & 0 & 0 & 0 & 0 & 0 & 0 \\
 0 & 0 & 0 & 0 & 0 & 1 & 0 & 0 & 0 & 0 & 0 & 0 & 0 & 0 \\
 0 & 1 & 0 & 0 & 0 & 0 & 0 & 0 & 0 & 0 & 96 & -55 & 2 & 4 \\
 0 & 0 & 1 & 0 & 0 & 0 & 0 & 0 & 0 & 0 & -96 & 55 & -2 & -4 \\
 0 & 0 & 0 & 0 & 0 & 0 & 0 & 0 & 1 & 0 & 0 & 0 & 0 & 0 \\
 0 & 0 & 0 & 0 & 0 & 0 & 0 & 0 & 0 & 1 & 0 & 0 & 0 & 0 \\
 0 & 0 & 0 & 0 & 0 & 0 & 0 & 0 & 0 & 0 & 1 & 0 & 0 & 0 \\
 0 & 0 & 0 & 0 & 0 & 0 & 0 & 0 & 0 & 0 & 0 & 1 & 0 & 0 \\
 0 & 0 & 0 & 0 & 0 & 0 & 0 & 0 & 0 & 0 & 0 & 0 & 1 & 0 \\
 0 & 0 & 0 & 0 & 0 & 0 & 0 & 0 & 0 & 0 & 0 & 0 & 0 & 1
\end{array}
\right)}}
\quad
\hat{M}_{{\rm D_{12}}}=
\mbox{\tiny\ensuremath{\arraycolsep=2pt
\left(
\begin{array}{cccccccccccccc}
 0 & 0 & 1 & 0 & 0 & 0 & 0 & 0 & 0 & 0 & -16 & 40 & 0 & -8 \\
 0 & 0 & 0 & 1 & 0 & 0 & 0 & 0 & 0 & 0 & -16 & 40 & 0 & -8 \\
 1 & 0 & 0 & 0 & 0 & 0 & 0 & 0 & 0 & 0 & 16 & -40 & 0 & 8 \\
 0 & 1 & 0 & 0 & 0 & 0 & 0 & 0 & 0 & 0 & 16 & -40 & 0 & 8 \\
 0 & 0 & 0 & 0 & 1 & 0 & 0 & 0 & 0 & 0 & 0 & 0 & 0 & 0 \\
 0 & 0 & 0 & 0 & 0 & 1 & 0 & 0 & 0 & 0 & 0 & 0 & 0 & 0 \\
 0 & 0 & 0 & 0 & 0 & 0 & 1 & 0 & 0 & 0 & 0 & 0 & 0 & 0 \\
 0 & 0 & 0 & 0 & 0 & 0 & 0 & 1 & 0 & 0 & 0 & 0 & 0 & 0 \\
 0 & 0 & 0 & 0 & 0 & 0 & 0 & 0 & 1 & 0 & 0 & 0 & 0 & 0 \\
 0 & 0 & 0 & 0 & 0 & 0 & 0 & 0 & 0 & 1 & 0 & 0 & 0 & 0 \\
 0 & 0 & 0 & 0 & 0 & 0 & 0 & 0 & 0 & 0 & 1 & 0 & 0 & 0 \\
 0 & 0 & 0 & 0 & 0 & 0 & 0 & 0 & 0 & 0 & 0 & 1 & 0 & 0 \\
 0 & 0 & 0 & 0 & 0 & 0 & 0 & 0 & 0 & 0 & 0 & 0 & 1 & 0 \\
 0 & 0 & 0 & 0 & 0 & 0 & 0 & 0 & 0 & 0 & 0 & 0 & 0 & 1
\end{array}
\right)}}
\end{equation}
\begin{equation}
\hat{M}_{{\rm D_{2}}}=
\mbox{\tiny\ensuremath{\arraycolsep=2pt
\left(
\begin{array}{cccccccccccccc}
 1 & 0 & 0 & 0 & 0 & 0 & 0 & 0 & 0 & 0 & 0 & 0 & 0 & 0 \\
 0 & 0 & 1 & 0 & 0 & 0 & 0 & 0 & 0 & 0 & -48 & 80 & 0 & 0 \\
 0 & 1 & 0 & 0 & 0 & 0 & 0 & 0 & 0 & 0 & 48 & -80 & 0 & 0 \\
 0 & 0 & 0 & 1 & 0 & 0 & 0 & 0 & 0 & 0 & 0 & 0 & 0 & 0 \\
 0 & 0 & 0 & 0 & 1 & 0 & 0 & 0 & 0 & 0 & 0 & 0 & 0 & 0 \\
 0 & 0 & 0 & 0 & 0 & 1 & 0 & 0 & 0 & 0 & 0 & 0 & 0 & 0 \\
 0 & 0 & 0 & 0 & 0 & 0 & 1 & 0 & 0 & 0 & 0 & 0 & 0 & 0 \\
 0 & 0 & 0 & 0 & 0 & 0 & 0 & 1 & 0 & 0 & 0 & 0 & 0 & 0 \\
 0 & 0 & 0 & 0 & 0 & 0 & 0 & 0 & 1 & 0 & 0 & 0 & 0 & 0 \\
 0 & 0 & 0 & 0 & 0 & 0 & 0 & 0 & 0 & 1 & 0 & 0 & 0 & 0 \\
 0 & 0 & 0 & 0 & 0 & 0 & 0 & 0 & 0 & 0 & 1 & 0 & 0 & 0 \\
 0 & 0 & 0 & 0 & 0 & 0 & 0 & 0 & 0 & 0 & 0 & 1 & 0 & 0 \\
 0 & 0 & 0 & 0 & 0 & 0 & 0 & 0 & 0 & 0 & 0 & 0 & 1 & 0 \\
 0 & 0 & 0 & 0 & 0 & 0 & 0 & 0 & 0 & 0 & 0 & 0 & 0 & 1
\end{array}
\right)}}
\end{equation}
By the ordering \eqref{namely}, the extended Gepner monodromy is given by
$$
\hat{M}_{\rm G}=  \hat{M}_{{\rm D}_{11}} \cdot \hat{M}_{\rm LV} \cdot \hat{M}_{{\rm D}_2}
\cdot \hat{M}_{\rm C} \cdot \hat{M}_{{\rm D}_{12}}
\,,
\qquad (\hat{M}_{\rm G})^5 = 1
$$
We also find the vanishing domain walls
\begin{equation}
\tau_7-\tau_2 -96\varpi_0+55\varpi_1-2\varpi_2-4\varpi_3\qquad
\text{and} \qquad
\tau_8-\tau_3+96 \varpi_0 - 55\varpi_1+2\varpi_2 +4\varpi_3
\end{equation}
at $\zDo{11}$,
\begin{equation}
\tau_1-\tau_3 +16\varpi_0-40\varpi_1+8\varpi_3
\qquad\text{and}\qquad
\tau_4-\tau_2 -16\varpi_0+40\varpi_1-8\varpi_3
\end{equation}
at $\zDo{12}$, and
\begin{equation}
\tau_3-\tau_2 -48\varpi_0+80\varpi_1
\end{equation}
at $\zDo2$.

\section{Discussion}
\label{discussion}

In this work, we have studied analytic invariants 
of a variety of (mostly new!) algebraic cycles on four
one-parameter Calabi-Yau hypersurfaces in weighted projective space. We found
these cycles by looking for holomorphic curves invariant under particular permutation
symmetries of the homogeneous coordinates. In section \ref{curves}, we calculated
the inhomogeneous Picard-Fuchs equation satisfied by the truncated normal function
associated with each cycle. In section \ref{algebraic}, we verified that
the large volume expansion satisfies the algebraic ``D-logarithm'' integrality
of \cite{arithmetic,svw}. In section \ref{algebraic}, we calculated the
monodromy representation \eqref{underlies} underlying the variation of mixed
Hodge structure. A by-product of these calculations is the limiting value 
of the normal function. We summarize our results in the table \ref{main}.
(For the van Geemen lines on the quintic, we have rescaled the results of 
\cite{lawa} by a factor of $4$ in order to conform to our present conventions.)

\renewcommand\arraystretch{1.6}
\newcolumntype{C}[1]{>{\centering\let\newline\\\arraybackslash\hspace{0pt}}p{#1}}
\newcolumntype{L}[1]{>{\raggedright\let\newline\\\arraybackslash\hspace{0pt}}m{#1}}
\newcolumntype{R}[1]{>{\raggedleft\let\newline\\\arraybackslash\hspace{0pt}}p{#1}}
\newcommand\fu{\newline\vskip -.0cm}

\begin{table}[t]
\begin{center}
\begin{tabular}{|l@{}l|l|c|l|c|}
\hline
 vernacular   &ref.\ & $r$ & $\zeta$ & $a$ &  \\
\hline\hline
van Geemen &\eqref{vG} & $1$ & $\sqrt{-3}$ &  $\ii 3.3421402589\ldots$ &  
$\displaystyle \frac{195\sqrt{-3}}{8\pi^2} \, L(2,\chi)$ \\[.1cm]
\hline
octic, first &\eqref{i1} & $4$ & $\sqrt{-3}$ & $\ii 3.0850525466\ldots$ & 
$\displaystyle \frac{45\sqrt{-3}}{2\pi^2} \, L(2,\chi)$ \\[.1cm]
\hline
---, second &\eqref{i2} & $1$ & $\sqrt{-7}$ & $\ii 6.4847457103\ldots$ & 
$\displaystyle \frac{21 \sqrt{-7}}{\pi^2} \, L(2,\chi)$ \\[.1cm]
\hline
dectic, first &\eqref{fgroup} & $2$ &   $\sqrt{-57}+\sqrt{-2}$ & $\ii 5.1547746324\ldots$ & 
$\displaystyle \frac{171\sqrt{-57}}{32\pi^2} \, L(2,\chi)$ \\[.1cm]
\hline
---, second &\eqref{collapses} & $3$ & $\sqrt{-3}$ & $\ii 5.0903367020\ldots$ & 
$\displaystyle \frac{297\sqrt{-3}}{8\pi^2} \, L(2,\chi)$ \\[.1cm]
\hline
quintic, real &\eqref{survives} & $1$ & $\sqrt{10(-5+\sqrt{85})}$ & $ 1.6377482973\ldots$ &  
?? \\ \hline
---, imaginary  &\eqref{survives} & $1$ & $\sqrt{10(-5-\sqrt{85})}$ & $\ii 93.620780658\ldots$ & 
?? \\\hline
---, second &\eqref{secondgroup} & $6$ & $5^{1/3}$ & $0$ & $0$
\\\hline
\end{tabular}
\caption{Arithmetic data of various algebraic cycles on Calabi-Yau hypersurfaces.}
\label{main}
\end{center}
\end{table}

The key formula to discuss the arithmetic data is the large volume
expansion \eqref{Amodel}. 
\begin{equation}
\eqlabel{focus}
\calw_A = \frac{s}{2\pi\ii r} \log q + a + \frac{1}{(2\pi\ii)^2}
\sum_{d=1}^\infty \tilde n_d q^{d/r}
\end{equation}
Namely, $r$ is the ramification (or ``Puiseux'') index of the corresponding 
branch of the covering \eqref{scheme}. The coefficients $\tilde n_d$ are algebraic 
numbers in a finite extension of $\rationals$, the ``residue field'', $K$. In our
terminology, we have referred to branches with the same $r$ and $K$ as ``a group
of branches''. The various branches in one group are distinguished by the
choice of phase of $q^{1/r}$, as well as the choice of an embedding of $K$ into 
$\complex$. [We might emphasize that there are cases in which the two choices are 
not independent. For instance, the ``second group'' of conics on the quintic 
\eqref{secondgroup} has $K=\rationals(5^{1/3})$, and $r=6$. But as the coefficients 
satisfy $\tilde n_d/5^{2d/3}\in\rationals$, only the overall choice of phase of
$5^{2/3} q^{1/6}$ matters, and there are really $6$ branches in that group. In
contrast, the first component of lines on the octic \eqref{upto}, has $r=4$
and $K=\rationals(\sqrt{-3})$. But the two choices are independent, and 
there are $8$ different branches. Somewhat more formally, we can have an
embedding of the Galois group of the local extension of moduli ($\cong\zet/r\zet$)
into the Galois group of the extension of residue field (or rather its Galois closure).
Another example of this is the first group on the dectic, eq.\ \eqref{fgroup}.]
The generator of this embedding is written as $\zeta$ in the table.
We emphasize the two most important aspects of the expansion \eqref{focus}.

\medskip

1. The coefficients $\tilde n_d$ satisfy ``Ooguri-Vafa integrality with an arithmetic twist''.
In the simplest cases, this means that there is a Dirichlet character $\chi$ such that in
\begin{equation}
\sum\tilde n_d q^{d/r} = \sum n_d \, {\rm Li}_2^{(\chi)} (q^{d/r})
\end{equation}
the $n_d$ are integral (at least outside the discriminant), where
\begin{equation}
{\rm Li}_2^{(\chi)}(q) = \sum \frac{\chi(k)}{k^2} q^k
\end{equation}
is the ``D-logarithm''. Specifically, for a quadratic extension $\zeta=\sqrt{\Delta}$,
we can write $\chi$ in terms of the Jacobi-symbol,
\begin{equation}
\eqlabel{char}
\chi(k) = \left(\frac{\Delta}{k}\right)
\end{equation}
In more complicated cases, the twist depends on $n_d$, see around eq.\ \eqref{around},
and \cite{arithmetic,svw}.

\smallskip

2. The constant term $a$ of the expansion is an interesting (conjecturally irrational)
number that we identify with the limiting value of the Abel-Jacobi map discussed in
\cite{ggk2}. [Comparison with section \ref{topological} will show that in many 
cases we have stripped off a simple rational additive that appears to be explained
by the branch structure at $\zLV$, rather than the intrinsic arithmetic of the
residue field.] As we have written in the table, in all cases with abelian Galois group, 
$a$ can be expressed in terms of the Dirichlet L-function with the
same character \eqref{char} that appears in the D-logarithm. We have not yet succeeded
in identifying an analogous formula in the non-abelian case (first group of conics
on the quintic). [The fact that $a$ vanishes (modulo $\rationals$) for the second 
group of conics on the quintic, and that the extension by $\sqrt{-2}$ disappears 
from $a$ in the first group on the dectic are consequences of the interplay with 
the extension of moduli at $\zLV$ that we mentioned above.]

\medskip

An interesting technical aspect of our calculations is that for each globally well-defined
cycle $\calc\to B$ (\cf, \eqref{scheme}), the number of independent limiting values 
that have to be calculated matches the number of components of the open string discriminant.
For instance, for the lines on the dectic, there are $2$ groups at $\zLV$, and $D$
has $2$ components. The conics on the quintic also split into two groups at $\zLV$.
However, because the residue field of the first group 
has two essentially independent embeddings into 
$\complex$ (corresponding to the real and imaginary $\zeta$), we really have three
independent values to calculate (counting one for $\rationals(5^{1/3})$). This is
matched precisely by the fact that the open string discriminant has three different 
components, see \eqref{shouldbe}. This state of affairs has allowed us to calculate
$a$ in two independent ways, from the conifold monodromy and the vanishing domain wall 
condition. We suspect that there is an underlying general statement.

\medskip

Considering which number fields appear in the examples, one might observe that lines
only come with abelian extensions of $\rationals$, though it is hard for us to tell 
whether this had to be true. The hunch that conics give at most solvable Galois
groups is dispelled by an example from \cite{arithmetic} (see table 1 there).

\medskip

Clearly the most interesting open problem is to find an A-model explanation for
the interesting arithmetic that we have observed here in the B-model. Some 
possibilities for the constant $a$ were mentioned in \cite{lawa}, and one
can be rather hopeful that one of them will materialize soon. For the 
D-logarithm integrality, we refer to the speculations in \cite{arithmetic}.

\begin{acknowledgments}
J.W.\ thanks the organizers of the workshop ``Recent advances in Hodge theory'',
University of British Columbia, June 2013, for the invitation to present some of 
these results.
This research is supported in part by an NSERC discovery grant, and a
Tier II Canada Research Chair.
\end{acknowledgments}

\newpage

\appendix

\section{Some Details of Residue Calculation}
\label{few}

For the convenience of the reader, we give a few details of the residue algorithm
developed in \cite{mowa,newissues}, applied to the first component of $\zet_3$-invariant
lines on the octic, see eq.\ \eqref{i1}.

We start from the expression eq.\ \eqref{convenient} for the holomorphic
three-form. With $z=\psi^{-8}$, the Picard-Fuchs operator \eqref{PF} has the form
\begin{equation}
\eqlabel{all}
\begin{split}
\call &= \theta^4 - z \textstyle(\theta+\frac 18)(\theta+\frac 38)(\theta+\frac 58)
(\theta+\frac 78) \\
&= \frac{1}{8^4 \psi^3} \Bigl(\textstyle{(\psi^8-1)\del_\psi^4 
+ \bigl(10 \psi^7+\frac{6}{\psi}\bigr) \del_\psi^3
+ \bigl(25\psi^6-\frac{15}{\psi^2}\bigr) \del_\psi^2
+ \bigl(15\psi^5+\frac{15}{\psi^3}\bigr)\del_\psi
+ \psi^4}\Bigr) \frac{1}{\psi}
\end{split}
\end{equation}
The Griffiths-Dwork reduction method now allows us to write the two-form in \eqref{where}
as
\begin{equation}
\beta = {\rm Res}\; \tilde\beta
\end{equation}
where $\tilde\beta$ is the meromorphic three-form\footnote{multiplied
with $8^4\psi^3$ and up to factors of $2\pi\ii$}
\begin{equation}
\eqlabel{most}
\begin{split}
\tilde \beta&=
\textstyle - \frac{6\omega_5 x_1^4 x_2^4 x_3^4 x_4^4 x_5^3}{W^4} 
- \frac{6\psi \omega_5 x_1^5 x_2^5 x_3^5 x_4^5 x_5^2}{W^4}
- \frac{6\psi^2 \omega_5 x_1^6 x_2^6 x_3^6 x_4^6 x_5}{W^4}
- \frac{6\psi^3 \omega_1 x_2^7 x_3^7 x_4^7 x_5}{W^4}
- \frac{6\psi^4\omega_2 x_2 x_3^8 x_4^8 x_5^2}{W^4} \\
&
\textstyle
- \frac{6\psi^5 \omega_3 x_1 x_2 x_3^2 x_4^9 x_5^3}{W^4}
- \frac{6\psi^6 \omega_4 x_1^2 x_2^2 x_3^2 x_4^3 x_5^4}{W^4}
- \frac{6\psi^7 \omega_5 x_1^3 x_2^3 x_3^3 x_4^3 x_5^4}{W^4}
+ \frac{6\omega_5 x_1^3 x_2^3 x_3^3 x_4^3 x_5^2}{\psi W^3}
+ \frac{4\omega_5 x_1^4 x_2^4 x_3^4 x_4^4 x_5}{W^3}
\\
&\textstyle
+ \frac{4\omega_5 x_1^3 x_2^3 x_3^3 x_4^3 x_5^2}{\psi W^3}
+ \frac{2\psi \omega_5 x_1^5 x_2^5 x_3^5 x_4^5}{W^3}
+ \frac{2\omega_5 x_1^4 x_2^4 x_3^4 x_4^4 x_5}{W^3}
+ \frac{2\omega_5 x_1^3 x_2^3 x_3^3 x_4^3 x_5^2}{\psi W^3}
- \frac{2\psi^4 \omega_3 x_3 x_4^8 x_5^2}{W^3}
\\
&\textstyle
- \frac{2\psi^5 \omega_4 x_1 x_2 x_3 x_4^2 x_5^3}{W^3}
- \frac{2\psi^6 \omega_5 x_1^2 x_2^2 x_3^2 x_4^2 x_5^3}{W^3}
- \frac{4\psi^5 \omega_4 x_1 x_2 x_3 x_4^2 x_5^3}{W^3}
- \frac{4\psi^6 \omega_5 x_1^2 x_2^2 x_3^2 x_4^2 x_5^3}{W^3}
- \frac{6\psi^6 \omega_5 x_1^2 x_2^2 x_3^2 x_4^2 x_5^3}{W^3}
\\
&\textstyle
- \frac{6\omega_5 x_1^2 x_2^2 x_3^2 x_4^2 x_5}{\psi^2 W^2}
- \frac{2\omega_5 x_1^3 x_2^3 x_3^3 x_4^3}{\psi W^2}
- \frac{2\omega_5 x_1^2 x_2^2 x_3^2 x_4^2 x_5}{\psi^2 W^2}
- \frac{4\omega_5 x_1^2 x_2^2 x_3^2 x_4^2 x_5}{\psi^2 W^2}
-  \frac{\omega_5 x_1^3 x_2^3 x_3^3 x_4^3}{\psi W^2}
\\
&\textstyle
-  \frac{\omega_5 x_1^2 x_2^2 x_3^2 x_4^2 x_5}{\psi^2 W^2}
- \frac{2\omega_5 x_1^2 x_2^2 x_3^2 x_4^2 x_5}{\psi^2 W^2}
-  \frac{\psi^4 \omega_4 x_4 x_5^2}{W^2}
-  \frac{\psi^5 \omega_5 x_1 x_2 x_3 x_4 x_5^2}{W^2}
- \frac{2\psi^5 \omega_5 x_1 x_2 x_3 x_4 x_5^2}{W^2}
\\
&\textstyle
- \frac{4\psi^5 \omega_5 x_1 x_2 x_3 x_4 x_5^2}{W^2}
+ \frac{6\omega_5 x_1 x_2 x_3 x_4}{\psi^3 W}
+ \frac{2\omega_5 x_1 x_2 x_3 x_4}{\psi^3 W}
+ \frac{4\omega_5 x_1 x_2 x_3 x_4}{\psi^3 W}
+  \frac{\omega_5 x_1 x_2 x_3 x_4}{\psi^3 W}
\\
&\textstyle
+ \frac{2\omega_5 x_1 x_2 x_3 x_4}{\psi^3 W}
-  \frac{\psi^4 \omega_5 x_5}{W}
\end{split}
\end{equation}
and for $i=1,\ldots 5$,
\begin{equation}
\eqlabel{threeforms}
\omega_i = \omega(\del_i)
\end{equation}
Now consider the line $C$ parameterized as in \eqref{nontrivial}, with
$a^2=3$, $c=b$, and $b^2-2ab\,\psi+21=0$. To calculate
the inhomogeneous Picard-Fuchs equation associated to $C$, we choose a three-chain 
$\Gamma$ with $\del\Gamma=C$, and then apply Griffiths' ``tube-over-cycle map''
to write
\begin{equation}
\eqlabel{ff}
\call\int^C\Omega = \call \int_{T_\epsilon(\Gamma)} \tilde \Omega
\end{equation}
The calculation in $\projective^4$ then splits in two types of contributions:
``the exact terms'',
\begin{equation}
\eqlabel{fexact}
f_{\rm exact} = \int_{T_\epsilon(C)} \tilde \beta
\end{equation}
and the ``direct terms'', $f_{\rm direct}$, which come from differentiating the tube 
over the 3-chain. If $n_\psi$ is the normal vector to $T_\epsilon(C)$ implementing
infinitesimal variation in $\psi$ direction, $f_{\rm direct}$ is obtained from
\eqref{all} by replacing $\del_\psi^k$ with
\begin{equation}
\eqlabel{feed}
\del_\psi^k \int_{T_{\epsilon}(\Gamma)} \tilde \Omega 
-\int_{T_{\epsilon}(\Gamma)} \del_\psi^k\tilde\Omega
= \sum_{j=0}^{k-1} \del_\psi^{k-1-j}\int_{T_\epsilon(C)}
(\del_\psi^j\tilde\Omega)(n_\psi)
\end{equation}
We emphasize that while the final result is well-defined and does not depend on any
choices (such as, the three-form $\tilde\beta$, or the tube $T_\epsilon(C)$), the 
decomposition into $f_{\rm exact}$ and $f_{\rm direct}$ in general will.

The freedom in laying the tube is the key to calculating the integrals 
defining $f_{\rm exact}$ and $f_{\rm direct}$ in terms of residues. Fix some choice 
of plane $P$ passing through $C$, and denote the residual curve by $R$,
\begin{equation}
\{W=0\} \cap P = C\cup R
\end{equation}
We can now lay the tube over $C$ inside of $P$ except for some neighborhood of
the intersection points
\begin{equation}
C\cap R = \{ p_1,\ldots ,p_s\}
\end{equation}
where we have to escape into the rest of $\projective^4$. A certain advantage of
the lines over the conics studied in \cite{mowa,arithmetic} is that we have 
some choice in picking $P$, whereas a conic already spans a plane. In the case at 
hand, it turns out convenient to let the second generator point in the $x_5$ 
direction, which results in the four intersection points
\begin{equation}
\eqlabel{four}
p_1 = \{u=0\}\,, \quad p_2=\{u+v =0\} \,, \quad p_{3,4}= \{u^2 - uv+v^2 =0\}
\end{equation}
Then, for each of those four points, we choose a third direction, normal to the plane,
and a local coordinate $z$ on the curve. We also need a real function $f(r)$ that 
smoothly decreases from $1$ to $0$ as $r=|z|$ runs from $0$ to some small positive 
$r_*$. The role of $f(r)$ is to return the tube to $P$ for $|z|>r_*$.

To be completely explicit, around $p_1$ above, with $z=u/v$, we escape into the 
$x_1$-direction. The tube is parameterized as
\begin{equation}
\begin{split}
x_1=1+z+ \epsilon\, r g(r) \,,\quad x_2= \omega+z  \,,\quad
x_3= \omega^2+z  \,,\quad x_4= a \, z \,,\\
x_5= c \, z + b \, z^4 + \epsilon \Bigl[\frac{r}{(b-a \psi) z}-\frac{7 r g(r)}{b-a \psi}\Bigr]
\qquad\qquad
\end{split}
\end{equation}
where for computational convenience, we have rewritten $f(r)$ as $rg(r)$.
Around $p_2$, with $z=u/v+1$, the $x_2$-direction turns out to be more convenient.
\begin{equation}
\begin{split}
x_1=z \,,\quad x_2=-1+\omega+z-\frac{\epsilon\, r g(r)}{27 (-1+\omega)} \,,
\quad
x_3= -1+\omega^2+z  \,,\quad x_4= a \,(-1+z) \\
x_5=c\, (-1+z)+b\, (-1+z)^4+
\epsilon\Bigl[\frac{-r}{3 (b-a \psi) z}+\frac{7 r g(r)}{3 (-1+\omega) (b-a \psi)}\Bigr]
\qquad\quad
\end{split}
\end{equation}
Finally, for $p_{3,4}$, we use $z=u/v-u_*$, where $u_*$ is one of the two roots
of $u^2-u+1=0$, and we again go in the $x_1$-direction
\begin{equation}
\begin{split}
x_1=1+u_*+z-\frac{\epsilon\, r g(r)}{27 (1+u_*)}\,,\;\;
x_2=\omega+u_*+z \,,\;\;
x_3=\omega^2+u_*+z \,, \;\;
x_4=a\,(u_*+z) \,,\\
x_5=c\, (u_*+z)+b\, (u_*+z)^4+\epsilon 
\Bigl[\frac{-r}{3 (b-a \psi) z}+\frac{7 r g(r)}{3 (b-a \psi) (1+u_*)}\Bigr]
\qquad\qquad\quad
\end{split}
\end{equation}
By construction, the restriction of $W$ to each of the tubes takes the form
\begin{equation}
\eqlabel{WT}
W|_{T_{\epsilon}(C)} \sim \epsilon \bigl(r+rg(r) + \calo(z^2)\bigr) + \calo(\epsilon^2)
\end{equation}
thereby exhibiting the order of pole of each of the terms in \eqref{threeforms}.
The convenience of choosing the plane in the $x_5$-direction becomes apparent 
when restricting the three-forms $\omega_i$ from \eqref{threeforms}. All terms in 
\eqref{most} involving $\omega_5$ vanish as a result of our choice.
As for the non-vanishing three-forms, we have around $p_1$, for example,
\begin{equation}
\begin{split}
\omega_2 &= \frac{a+a \omega}{b-a\psi} \,z^{-1} \epsilon r^2  g'(r) dz d\epsilon dr \\
\omega_3 &= \frac{a\, \omega}{b-a \psi} \,z^{-1} \epsilon r^2  g'(r) dz d\epsilon dr \\
\omega_4 &= \frac{-1-2 \omega}{b-a\psi}\,z^{-1} \epsilon r^2  g'(r) dz d\epsilon dr
\end{split}
\end{equation} 
The algebraic calculation of the residues is then accomplished by expanding \eqref{WT}
to the order in $\epsilon$ and $z$ dictated by the pole order of the three-form 
under consideration, picking out the term of degree $0$, and integrating over
$r$. We do this around each of the four points \eqref{four}, and sum up the results.

For $f_{\rm exact}$, we find, corresponding the $37$ terms in \eqref{threeforms}
\begin{multline}
0+0+0+
\textstyle{\frac{2 (-31 a b \psi^4-62 a b \omega \psi^4+93 \psi^5+186 \omega 
\psi^5-5 a b \psi^6-10 a b \omega \psi^6+15 \psi^7+30 \omega \psi^7)}{3 (-7+\psi^2)^3}} \\
+
\textstyle{\frac{5 (-7 a b \psi^4-14 a b \omega \psi^4+21 \psi^5+42 \omega \psi^5-2 a b 
\psi^6-4 a b \omega \psi^6+6 \psi^7+12 \omega \psi^7)}{3 (-7+\psi^2)^3}}
-\frac{7 (-a b \psi^6-2 a b \omega \psi^6+3 \psi^7+6 \omega \psi^7)}{(-7+\psi^2)^3}\\
+0+0+0+0+0+0+0+0+
\textstyle{\frac{7 (-a b \psi^4-2 a b \omega \psi^4+3 \psi^5+6 \omega \psi^5)}{3 (-7+\psi^2)^2}}
+0+0+\cdots
\end{multline}
For $f_{\rm direct}$, we first have to calculate $n_\psi$ by differentiating the 
parameters entering the tube (which depend implicitly on $\psi$). We then
contract $n_\psi$ with $\del_\psi^j\tilde\Omega$ and feed the result into \eqref{feed}.
In the case at hand, $f_{\rm exact}$ turns out to vanish. Remembering some overall 
factors, and simplifying judiciously, the final result becomes precisely \eqref{f1}.


\end{document}